\documentclass[twocolumn]{svjour3}

\smartqed

\usepackage{amsmath,amssymb,epsfig}
\usepackage{graphicx}
\usepackage{natbib}

\newcommand{\thm}{\begin{theorem}}
\newcommand{\lem}{\begin{lemma}}
\newcommand{\pro}{\begin{proposition}}
\newcommand{\dfn}{\begin{definition} \rm}
\newcommand{\rem}{\begin{remark}}
\newcommand{\xam}{\begin{example}}
\newcommand{\cor}{\begin{corollary}}
\newcommand{\prf}{\begin{proof}}
\newcommand{\ethm}{\end{theorem}}
\newcommand{\elem}{\end{lemma}}
\newcommand{\epro}{\end{proposition}}
\newcommand{\edfn}{\end{definition}}
\newcommand{\erem}{\bbox\end{remark}}
\newcommand{\exam}{\bbox\end{example}}
\newcommand{\ecor}{\end{corollary}}
\newcommand{\eprf}{\bbox \end{proof}}
\newcommand{\beqn}{\begin{equation}}
\newcommand{\eeqn}{\end{equation}}

\newcommand{\bbox}{\vrule height7pt width4pt depth1pt}
\newcommand{\commentout}[1]{}
\newcommand{\M}{{\cal M}}

\newcommand{\BR}{\mathit{BR}}

\newenvironment{RETHM}[2]{\trivlist \item[\hskip 10pt\hskip\labelsep{\sc #1\hskip 5pt\relax\ref{#2}.}]\it}{\endtrivlist}
\newcommand{\rethm}[1]{\begin{RETHM}{Theorem}{#1}}
\newcommand{\repro}[1]{\begin{RETHM}{Proposition}{#1}}
\newcommand{\relem}[1]{\begin{RETHM}{Lemma}{#1}}
\newcommand{\recor}[1]{\begin{RETHM}{Corollary}{#1}}

\newcommand{\erethm}{\end{RETHM}}
\newcommand{\erepro}{\end{RETHM}}
\newcommand{\erelem}{\end{RETHM}}
\newcommand{\erecor}{\end{RETHM}}

\begin{document}

\title{Optimizing Scrip Systems: Crashes, Altruists, Hoarders, Sybils
  and Collusion\thanks{%
Portion of the material in this paper appeared in preliminary
form in papers
in the Proceedings of the 7th and 8th ACM Conferences on Electronic
Commerce~\cite{scrip06,scrip07} and the Proceedings of the First
Conference on Auctions, Market Mechanisms and Their
Applications~\cite{scrip09}.
Section~\ref{sec:model} and Appendix~\ref{sec:MDP} are reproduced from
a companion paper~\cite{scripjournal1} to make this paper
self-contained.
Much of the work was performed while Ian Kash and Eric Friedman were at Cornell University.}}

\author{Ian A. Kash \and
        Eric J. Friedman \and
        Joseph Y. Halpern}

\institute{I. A. Kash \at
                Microsoft Research Cambridge, 7 J. J. Thomson Ave, Cambridge, UK CB3 0FB\\
                 \email{iankash@microsoft.com}
           \and
           E. J. Friedman \at
                 Internation Computer Science Institute and Department of Computer Science, University of California, Soda Hall, Berkeley, CA 94720\\
                 \email{ejf@icsi.berkeley.edu}
           \and
           J. Y. Halpern \at
                 Computer Science Department, Cornell University,
                 Upson Hall, Ithaca, NY 14850\\
                 \email{halpern@cs.cornell.edu}
}

\date{Received:  03-09-2010 / Accepted: 02-09-2011}

\maketitle

\begin{abstract}
Scrip, or artificial currency, is a useful tool for designing systems
that are robust to selfish behavior by users.  However, it also
introduces problems for a system designer, such as how the amount of
money in the system should be set.  In this paper,
the effect of varying the total amount of money
in a scrip system on
efficiency (i.e., social welfare---the total utility of
all the agents in the system) is analyzed, and it
is shown that by
maintaining the appropriate ratio between the total amount of money
and the number of agents, efficiency is maximized.
This ratio can be found
by increasing the money supply to just below the point that the system
would experience
a ``monetary crash,'' where money is sufficiently
devalued that no agent is willing to perform a service.
The implications of the presence of altruists, hoarders, sybils, and
collusion on the performance of the system are examined.
Approaches are discussed to identify the strategies and types of
agents.
\end{abstract}






\section{Introduction}\label{sec:intro}

Money is a powerful tool for dealing with selfish behavior.  For
example, in peer-to-peer systems, a common problem is \emph{free
riding}~\cite{adar00}, where users take advantage of the resources
offered by a system without contributing their own.
One way of dealing with the problem is to have users pay
for the use of others' resources, and to pay them for contributing their
own resources.
The incentive to free ride then disappears.  Similarly, monetary
incentives are a potential solution to resource allocation problems in
distributed and peer-to-peer systems: this is the business
model of cloud computing services such as Amazon EC2.

In some systems, it may not be desirable to use actual money.  As a
result, many systems have used an artificial currency,
or \emph{scrip}. (See
\cite{gupta03,fileteller02,agora,yootles,mariposa,karma03,egg,antfarm,aperjis08}
for some examples of the use of scrip in systems.)
While using scrip avoids some issues, such as processing payments, it
introduces new questions a system designer must face.  How much money
should be printed?  What should happen if the system grows rapidly or
there is significant churn?  What will happen if a small number of
users start hoarding money or creating sybils?

\commentout{
Historically, non-governmental organizations have issued their own
currencies, known as \emph{scrip}, for a wide variety of purposes.
Creating a scrip system
creates a new market where goods and services
can be exchanged that may have been impractical or undesirable to
implement with standard currency.
As a result, such currencies, known as \emph{scrip}, have been
used in many societies.   More recently, they have been used
in distributed and peer-to-peer systems, both to prevent
what is known as \emph{free riding}, where agents
take advantage of the resources the system provides while providing none
of their own (as Adar and Huberman \citeyear{adar00} have shown, this
behavior certainly takes place in systems such as Gnutella), and for
resource allocation.  (See
\cite{gupta03,fileteller02,agora,yootles,mariposa,karma03} for some
examples of the use of scrip in systems.)
}

The
story of the Capitol Hill Baby Sitting Co-op \cite{babysitting},
popularized by Krugman \citeyear{Krugman}, provides a cautionary tale
of how system performance can suffer if these issues are not handled
appropriately.
The Capitol Hill Baby Sitting Co-op was a group of parents working on
Capitol Hill who agreed to cooperate to provide babysitting services
to each other.  In order to enforce fairness, they issued a supply of
scrip with each coupon worth a half hour of babysitting.  At one
point, the co-op had a recession.  Many people wanted to save up
coupons for when they wanted to spend an evening out.  As a result,
they went out less and looked for more opportunities to babysit.
Since a couple could earn coupons only when another couple went out,
no one could accumulate more, and the problem only got worse.

After a
number of failed attempts to solve the problem, such as mandating a
certain frequency of
going out, the co-op started issuing more coupons.  The results were striking.
Since couples had a sufficient reserve of coupons, they were more
comfortable spending them.  This in turn made it much easier to earn
coupons when a couple's supply got low.
Unfortunately, the amount of scrip grew to the point
that most of the couples felt ``rich.''  They had
enough scrip for the foreseeable future, so naturally they didn't want
to devote their evening to babysitting.  As a result,
couples who wanted to go out were unable to find another couple willing
to babysit.

This anecdote shows that the amount of scrip in circulation can
have a significant impact on the effectiveness of a scrip
system.
If there is too little money in the system, few agents will be
able to afford service.
At the other extreme, if
there is too much money in the system, people
feel rich and stop looking for work. Both of these extremes lead to
inefficient outcomes.  This suggests that there is an optimal amount
of money, and that
nontrivial deviations from the optimum towards either extreme
can lead to significant degradation in the performance of the system.

In a companion paper \cite{scripjournal1}, we gave a formal model of a
scrip system and studied the behavior of scrip systems from a
micro-economic, game-theoretic viewpoint.  Roughly speaking, in the
model agents want work done at random times.  To get the work done,
they must have at least \$1 of scrip.%
\footnote{Although we refer to our unit of scrip as the dollar,
these are not real dollars, nor do we view ``scrip dollars'' as
convertible to real dollars.}  All the agents willing to do the work
volunteer to do it, and one is chosen at random (although not
necessarily uniformly at random; agents may have different likelihoods
of being chosen).  The agent who has the work done pays \$1 to the agent
chosen to do it, and gains 1 unit of utility, while the agent who does
the work suffers a small utility loss.   We showed that in such a scrip
system, there is a nontrivial Nash equilibrium where all agents use a
\emph{threshold strategy}---that is, agent $i$ volunteers to work iff
$i$ has below some threshold of $k_i$ dollars (the threshold may be
different for different agents).  A key part of our analysis
involves a characterization of the distribution of wealth when agents
all use threshold strategies.

In this paper, we use the analysis of scrip systems from
~\cite{scripjournal1} to understand how robust
scrip systems are, and how to optimize their performance.
We compute the money supply
that maximizes social welfare, given the number
of agents.
As we show, the behavior mimics the behavior in the babysitting coop
example.  Specifically, if we start with a system with relatively little
money (where ``relatively little'' is measured in terms of the average
amount of money per agent), adding more money decreases the number of
agents with no money, and thus increasing social welfare.  (Since it is
more likely that an agent will be able to pay for someone to work when
he wants a job done.)
However, this only works up to a point.
Once
a critical
amount of money is reached, the system experiences a monetary crash:
just as in the babysitting coop example,
there is so much money that, in equilibrium, everyone will feel rich and
no agents are willing to work.
We show that,
to get optimal performance, we want the total amount of money in the
system to be as close as possible to the critical amount, but not to go over
it.  If the amount of money in the system is over the critical amount,
we get the worst possible equilibrium, where
no agent ever volunteers, and
all agents have utility 0.
This means that, for a system designers' point of view, there is a
significant tradeoff between
efficiency and robustness.

The equilibrium analysis in~\cite{scripjournal1}
assumes that all agents have somewhat similar
motivation: in particular, they do not get pleasure simply from
performing a service, and are interested in money only to the extent
that they can use it to get services performed.
But in real systems, not all agents
have this motivation.
Some of the more common ``nonstandard'' agents are \emph{altruists} and
\emph{hoarders}.
Altruists are willing to satisfy all requests,
even if they go unpaid (think of babysitters who love kids so much that
they get pleasure from babysitting, and are willing to babysit for
free); hoarders value scrip for its own sake and are willing to
accumulate amounts far beyond what is actually useful.
Studies of the Gnutella peer-to-peer file-sharing network have shown
that one percent of agents satisfy fifty percent of the
requests~\cite{adar00,Hughes05}.
These agents are doing significantly more work for
others than they will ever have done for them,
so can be viewed as altruists.
Anecdotal evidence also suggests that the introduction of any sort of
currency seems to inspire hoarding behavior on the part of some agents,
regardless of the benefit of possessing money.
For example, SETI@home has found that contributors put in significant
effort to make it to the top of their contributor rankings.  This has
included returning fake results of computations rather than performing
them~\cite{zhao05}.

Altruists and hoarders have opposite effects on a system: having
altruists has the same effect as adding money;
having hoarders
is essentially equivalent to removing it.
With enough altruists in the system, the system has a monetary crash,
in the sense
that standard agents stop being willing
to provide service,
just as when there is too much money in the system
(the system still functions on a limited basis after the monetary
crash because altruists still supply service).
We show that, until we get to the point where the
system crashes,
the utility of
standard agents is improved by the presence of altruists.
We show that the presence of altruists makes the critical point lower
than it  would without them.  Thus, a system designer trying to optimize
the performance of the system by making the money supply as close as
possible to the critical point (but under it, since being over it would
result in a ``crash'') needs to be careful about estimating the number of
altruists.

Similarly, we show that, as the fraction of
hoarders
increases, standard agents begin to suffer because there is
effectively less money in circulation.
On the other hand, hoarders can improve the
stability of a system.  Since hoarders are willing to accept an
infinite amount of money, they can
prevent the monetary crash that would
otherwise occur as money was added.
In any case, our results show how the presence of both altruists and
hoarders can be
mitigated by appropriately controlling the money supply.

Beyond nonstandard agents, we also consider two different
manipulative behaviors in
which standard agents may engage: creating multiple identities,
known as \emph{sybils} \cite{sybil},
and collusion.
In scrip systems where each new user is given an initial amount
of scrip, there is an obvious benefit to creating sybils.
Even if this incentive is removed, sybils are still
useful: they can be used to increase the likelihood that an agent will be
asked to provide service, which makes it easier for him to earn
money.    This increases the utility of the sybilling agent, at the
expense of other agents,
in a manner reminiscent of the large view attack on
BitTorrent~\cite{largeview}.
From the perspective of an agent considering creating sybils,
the first few sybils can provide him with a significant benefit, but
the benefits of additional sybils
rapidly diminish.  So if a designer can make sybilling moderately
costly, the number of sybils actually created by rational agents will
usually be relatively small.

If a small fraction of agents have sybils, the situation is more
subtle.  Agents with sybils still do better than those without,
but the situation is not zero-sum.  In particular,
even agents without sybils might be better off, due to having more
opportunities to earn money.   Somewhat surprisingly, sybils can
actually result in everyone being better off.
However, exploiting this fact is generally not desirable.
The same process that leads to an
improvement in social welfare can also lead to a monetary crash,
where all agents stop providing service.
The system designer can
achieve the same effects  by increasing the average amount of
money or biasing the volunteer selection process.  In practice, it
seems better to do this than to exploit the possibility of sybils.

In our setting, having sybils is helpful because it increases the
likelihood that an agent will
be asked to provide service.  Our analysis of sybils applies no matter
how this increase in likelihood occurs.  In particular,
it applies to advertising.  Thus, our results suggest that there are
tradeoffs involved in allowing advertising.
For example, many systems allow agents to announce their connection
speed and other similar factors.  If this biases requests towards
agents with high connection speeds, even when agents with lower
connection speeds are perfectly capable of satisfying a particular
request, then agents with low connection speeds will have
an unnecessarily worsened
experience in the system.  This also means that
such agents will have a strong incentive to lie about their connection
speed.

While collusion is considered a bad thing in most systems, in the context of
scrip systems with fixed prices, it is almost entirely positive.
Without collusion, if
a user runs out of money he is unable to request service until he is
able to earn some.
However, a colluding group can pool there money so that all members
can make a request whenever the group as a whole has some money.
This increases
welfare for the agents who collude because
agents who have no money receive no service.
Collusion tends to benefit the non-colluding agents
as well.  Since colluding agents work less often,
it is easier for everyone to earn money, which ends up making everyone
better off.  However, as with sybils, collusion does
have the potential of crashing the system if the average amount of
money is too close to the critical point.

While a designer should generally encourage collusion in scrip systems,
we would expect that in most systems there will be relatively little
collusion, and
what collusion exists will involve small numbers of agents.  After
all, scrip systems exist to try and resolve resource-allocation
problems.  If agents could
collude to optimally allocate resources within the group, they would
not need a scrip system in the first place.
Nevertheless, our analysis of collusion indicates a way that system
performance could be improved even without collusion.  Many of the
benefits of collusion come from agents effectively being allowed to
have a negative amount of money (by borrowing from their
the other agents with whom they are colluding).  These benefits could
also be realized if agents
are allowed
to borrow money, so designing a loan mechanism could be an important
improvement for a scrip system.  Of course, implementing such a loan
mechanism in a way that prevents abuse requires a careful design.

\commentout{
The effects of altruists, sybils, and collusion on system
behavior have all been studied in other contexts.
Work on the evolution of cooperation stresses the importance of
altruists willing to undertake costly punishment~\cite{Nowak07}.
Yokoo et al.~\citeyear{yokoo04} studied the effects of sybils in
auctions.
Solution concepts such as \emph{strong Nash equilibrium}~\cite{strong}
and \emph{$k$-$t$ robust equilibrium}~\cite{ADGH06} explicitly address
collusion in games; Hayrapetyan et al.~\citeyear{hayrapetyan06} study
collusion in
congestion games and find cases where, as with scrip systems,
collusion is actually beneficial.
}

The analysis we carry out here has a benefit beyond showing
how to deal with altruists, hoarders, sybils, and collusion.
In order to utilize our results effectively, a system designer needs to
be able to identify characteristics of agents (with what frequency do
they make requests, how likely are they to be chosen to satisfy a
request, and so on) and what strategies they are following.  This is
particularly useful because finding an amount of money close to the point of
monetary crash, but not past it, relies on an understanding of the
agents in the system.  Of course, such information is also of great
interest to social scientists and marketers.
We show how relatively simple observations of the system can be used
to infer this information.

The rest of the paper is organized as follows.
In Section~\ref{sec:model}, we repeat the formal model
from~\cite{scripjournal1}.
Then in Section~\ref{sec:summary}, we summarize the results from that
paper.
We begin applying these results
in Section~\ref{sec:optimize}, where we show that the analysis leads
to an understanding of how to choose the amount of money in the system
(or, equivalently, the cost to fulfill a request) so as to maximize
efficiency, and also shows how to handle new users.
In Section~\ref{sec:nonstandard}, we discuss how the model can be used
to understand the effects of altruists, hoarders, sybils, and
collusion and provide guidance about how system designers can handle
these user behaviors.
All of this guidance relies on being able to understand what
strategies agents are using and what their preferences are.  In
Section~\ref{sec:identify}, we discuss how these can be inferred by
examining the system.
We conclude in Section~\ref{sec:conclusion}.

\section{The Model}\label{sec:model}

For the convenience of the reader we repeat Section 3 of our companion
paper which describes the model~\cite{scripjournal1}.

Before specifying our model formally, we give an intuitive description
of what our model captures.
While our model simplifies a number of features (as does any model),
we believe that it provides useful insights.
We model a scrip system where, as in a
P2P filesharing system, agents provide each other with service.  There is a
single service (such as file uploading) that agents occasionally want.  In
practice, at any given time, a number of agents will want service but, to
simplify the formal description and analysis, we model the scrip system
as proceeding in a series of rounds where, in each round, a single agent
wants service (the time between rounds will be adjusted to capture the
growth in parallelism as the number of agents grows).%
\footnote{For large numbers of agents, our model converges to one in
which agents
make requests in real time, and the time between an agent's requests
is exponentially distributed.  In addition, the time between requests
served by a single player is also exponentially distributed.}
In each round,
after an agent requests service, other agents have to decide whether
they want to volunteer to provide service.  However, not all agents
may be able to satisfy the request (not everyone has every file).
While, in practice, the ability of agents to provide service at
various times may be correlated for a number of reasons (if I don't
have the file today I probably still don't have it tomorrow; if one
agent does not have a file, it may be because it is rare, so that
should increase the probability that other agents do not have it), for
simplicity, we assume that the events of an agent being able to
provide service in different rounds or two agents being able to provide
service in the same or different rounds are independent.
While our analysis relies on this assumption so that we can describe
the system using a Markov chain, we expect that our results would
still hold as long these correlations are sufficiently small.
If there is at least one volunteer, someone is chosen from among the
volunteers
(at random)
to satisfy the
request.
Our model allows some agents to be more likely to be chosen (perhaps
they have more bandwidth) but
does not allow an agent to specify which agent is chosen.
Allowing agents this type of control would break the symmetries we use
to characterize the long run behavior of the system and create new
opportunities for strategic behavior by agents that are beyond the
scope of this paper.
The requester then gains some utility (he got the file) and
the volunteer loses some utility (he had to use his bandwidth to
upload the file), and the requester pays the volunteer a fee that
we fix at one dollar.  As is
standard in the literature, we assume that agents discount future
payoffs. This captures the
intuition that a util now is worth more than a util tomorrow, and
allows us to compute the total utility derived by an agent in an
infinite game.
The amount of utility gained by having a service performed and the
amount lost be performing it, as well as many other parameters may
depend on the agent.

More formally, we assume that agents have a \emph{type} $t$ drawn from
some finite set $T$ of types.
We can describe the entire population of agents
using the pair
$(T,\vec{f})$, where $\vec{f}$ is a vector of length $|T|$ and
$f_t$ is the fraction with type $t$.
For most of the paper, we consider only what we call \emph{standard
agents}.  These are agents who derive no pleasure from performing a
service, and for whom money has no intrinsic value.
Thus, for a standard agent, there is no direct connection between money
(measured in dollars) and utility (measured in utils).
We can characterize
the type of a standard agent by a tuple
$t = (\alpha_t, \beta_t, \gamma_t, \delta_t, \rho_t, \chi_t)$, where
\begin{itemize}
\item $\alpha_t > 0$ reflects the cost for an agent of type $t$ to
satisfy a request;
\item $0 < \beta_t < 1$ is the probability that an agent of type $t$ can
satisfy a request;
\item $\gamma_t > \alpha_t$ is the utility that an agent
of type $t$
gains for having a request satisfied;
\item $0 < \delta_t < 1$ is the rate at which an agent of type $t$ discounts
utility;
\item $\rho_t > 0$ represents the (relative) request rate (some people
want files more often than others).
For example, if there are
two types of agents with $\rho_{t_1} = 2$ and $\rho_{t_2} = 1$
then agents of the first type will make requests twice as often as
agents of the second type.  Since these request rates are relative, we
can multiply them all by a constant to normalize them.  To simplify
later notation, we assume the $\rho_t$ are normalized so that
$\sum_{t \in T} \rho_t f_t = 1$; and
\item $\chi_t > 0$ represents the (relative) likelihood of an agent to be
chosen when he volunteers (some uploaders may be more popular than
others).
In particular, this means the relative probability of two given agents
being chosen is independent of which other agents
volunteer.
\item $\omega_t = \beta_t \chi_t / \rho_t$ is not part of the tuple,
but is an important derived parameter that helps determine how much money an
agent will have.
\end{itemize}
We occasionally omit the subscript $t$ on some of these parameters when
it is clear from context or irrelevant.

Representing the population of agents in a system as $(T,\vec{f})$ captures
the essential features of a scrip system we want to model: there are a
large number of agents who may have different types.
Note that fixing a particular tuple $(T,\vec{f})$ puts a constraint on
the number $N$ of agents.
For example, if there are two types, and
$\vec{f}$ says that half of the agents are of each type, then there
cannot be 101 agents.
Similar issues arise when we want to talk about the amount of money in
a system.  We could deal with this problem in a number of ways (for example,
by having each agent determine his type at random according to the
distribution $\vec{f}$).
For convenience, we
simply do not consider population sizes that are incompatible with $\vec{f}$.
This is the approach used in the literature on
\emph{$N$-replica economies}~\cite{mascolell}.

Formally, we consider games specified by a tuple
$(T,\vec{f},h,m,n)$, where $T$ and $\vec{f}$ are as defined above,
$h \in \mathbb{N}$ is the \emph{base} number of agents of each type,
$n \in \mathbb{N}$ is number of \emph{replicas} of these agents and
$m \in \mathbb{R}^+$ is the
average amount of money.  The total number of agents is thus $hn$.  We
ensure that the
fraction
of agents of type $t$ is exactly $f_t$ and that
the average amount of money is exactly $m$ by requiring that $f_th \in
\mathbb{N}$ and $mh \in \mathbb{N}$.  Having created a base population
satisfying these constraints, we can make an arbitrary number of copies
of it.  More precisely,
we assume that
agents $0 \ldots f_{t_1}h - 1$ have
type $t_1$, agents $f_{t_1}h \ldots (f_{t_1} + f_{t_2})h - 1$ have type
$t_2$, and so on through agent $h - 1$.
These base agents determine the types of all other agents.
Each agent $j \in \{h, \ldots, hn-1\}$ has the same type as $j \mod
h$; that is, all the agents of the form $j + kh$ for $k = 1, \ldots,
n-1$ are replicas of agent $j$.

We also need to specify how money is initially allocated to agents.
Our results are based on the long-run behavior of the system and so
they turn out to hold for any initial allocation of money.  For
simplicity, at the start of the game we
allocate each of the $hmn$ dollars in the system
to an agent chosen uniformly at random,
but all our results would hold if we chose any other initial
distribution of money.

To make precise our earlier informal description,
we describe $(T,\vec{f},h,m,n)$ as an infinite extensive-form game.
A non-root node in the game
tree is associated with a round number (how many requests have
been made so far), a phase number, either 1, 2, 3 , or 4
(which describes how far along we are in determining the results of
the current request),
a vector $\vec{x}$ where $x_i$ is
the current amount of money agent $i$ has, and $\sum_i x_i = mhn$, and,
for some nodes, some
additional information whose role will be made clear below.
We use $\tau(i)$ to
denote the type of agent $i$.

\begin{itemize}

\item The game starts at a special root node, denoted $\Lambda$, where
nature moves.  Intuitively, at $\Lambda$, nature allocates money
uniformly at random, so it transitions to a node of the form
$(0,1,\vec{x})$: round
zero, phase one, and allocation of money $\vec{x}$, and  each possible
transition is
equally likely.

\item At a node of the form $(r,1,\vec{x})$, nature selects an agent
to make a request in the current round.  Agent $i$ is chosen with
probability $\rho_{\tau(i)} / hn$.
(Note that this is a probability because
$\sum_i \rho_{\tau(i)} = \sum_t f_t h n \rho_t = hn$.)
If $i$ is chosen, a transition is
made to
$(r,2,\vec{x},i)$.

\item At a node of the form $(r,2,\vec{x},i)$, nature selects the set
$V$ of agents (not including $i$) able to satisfy the request.  Each
agent $j \neq i$ is
included in $V$ with probability $\beta_{\tau(j)}$.  If $V$ is chosen, a
transition is made to $(r,3,\vec{x},i,V)$.

\item At a node of the form $(r,3,\vec{x},i,V)$, each agent in $V$
chooses whether to volunteer.  If $V'$ is the set of agents who choose
to volunteer, a transition is made to $(r,4,\vec{x},i,V')$.

\item At a node of the form $(r,4,\vec{x},i,V')$,
if $V' \neq \emptyset$,
nature chooses a single agent in $V'$ to satisfy the request.
Each agent $j$ is chosen with probability
$\chi_{\tau(j)} / \sum_{j' \in V'} \chi_{\tau(j')}$.  If $j$ is chosen,
a transition is made to $(r+1,1,\vec{x}')$, where
$$x_{\ell}' = \left \{
\begin{array}{lll}
x_{\ell} - 1 & \mbox{if } \ell = i \mbox{ and } x_i > 0,\\
x_{\ell} + 1 & \mbox{if } \ell \mbox{ is chosen by nature and } x_i > 0,\\
x_{\ell} & \mbox{otherwise.}\\
\end{array} \right.$$
If $V' = \emptyset$
or $x_i = 0$,
nature has no choice; a transition is made to
$(r+1,1,\vec{x})$ with probability 1.
\end{itemize}

A strategy for agent $j$ describes whether or not he will
volunteer at every node of the form $(r,3,\vec{x},i,V)$ such that $j \in
V$.  (These are the only nodes where $j$ can move.)
We also need to specify what agents know when they make their
decisions.  To make our results as strong as possible, we allow an
agent to base his strategy on the entire history of the game,
which includes, for example, the current wealth of every other agent.
As we show, even with this unrealistic amount of information,
available, it would still be approximately optimal to adopt a simple
strategy that requires little information---specifically, agents need to
know only their current wealth.  That means that our results would
continue to hold as long as agents knew at least this information.
A strategy profile $\vec{S}$ consists of one strategy per agent.
A strategy profile $\vec{S}$ determines a probability
distribution over paths $\Pr_{\vec{S}}$ in the game tree.  Each path
determines the
value of the following random two variables:

\begin{itemize}

\item $x_i^r$, the amount of money agent $i$ has during round $r$,
defined as the value of $x_i$ at the nodes with round number $r$ and

\item $u_i^r$, the utility of agent $i$ for round $r$.
If $i$ is a standard agent, then
$$u_i^r = \left \{
\begin{array}{lll}
\gamma_{\tau(i)} & \mbox{if the path has a node } (r,4,\vec{x},i,V' \neq 0)\\
-\alpha_{\tau(i)} & \mbox{if } i \mbox{ is chosen at node }
(r,4,\vec{x},j,V')\\
0 & \mbox{otherwise.}\\
\end{array} \right.$$

\end{itemize}


$U_i(\vec{S})$, the total expected utility of agent $i$ if strategy
 profile $\vec{S}$ is played, is the discounted sum of his
per round utilities $u_i^r$, but the exact form
of the discounting requires some explanation.  In our model, only one
agent makes a request each round.  As the number of agents increases,
an agent has to wait a larger number of rounds to make requests, so
naively discounting utility would mean his utility decreases as the
number of agents increases, even if all of his requests are
satisfied.
This is an artifact our model breaking time into discrete rounds where
a single agent makes a request.  In reality, many agents make requests
in parallel, and how often an agent desires service typically does not
depend on the number of agents.  It would be counterintuitive to have
a model that says that if agents make requests at a fixed rate and
they are all satisfied, then their expected utility depends on the
number of other agents.  As the following lemma shows, there is a
unique discount rate that removes this dependency.%
\footnote{In preliminary versions of this work we used the discount
  rate of $\delta_t^{1/n}$.  This rate captures the intuitive idea of making
  the time between rounds $1/n$, but results in an agent's utility
  depending on the number of other agents, even if all the agent's
  requests are satisfied.
  However, in the limit as $\delta_t$ goes to 1, agents' normalized
  expected utilities (multiplied by $1 - \delta_t$ as in
  Equation~\ref{eqn:U}) are the same either discount rate, so our
  main results hold with the discount rate  $\delta_t^{1/n}$ as well.}

\lem
With a discount rate of $(1 - (1-\delta_t)/n)$, an agent of type
$t$'s expected discounted utility for having all his requests
satisfied is independent of the number of replicas $n$.  Furthermore,
this is the unique such rate such that the discount rate is
$\delta_t$ when $n=1$.
\elem
\prf
The agent makes a request each round with probability $\rho_t / hn$,
so his expected discounted utility for having all his requests
satisfied is
\begin{align*}
&\sum_{r = 0}^{\infty} (1 - (1-\delta_t)/n)^r
(\rho_t \gamma_t / (hn))\\
&= (\rho_t \gamma_t / (hn)) / (1 - (1 - (1 - \delta_t)/n))\\
&= (\rho_t \gamma_t / h) / (1 - \delta_t)
\end{align*}
This is independent of $n$
and satisfies
$(1 - (1 - \delta_t)/1) = \delta_t$ as desired.
The choice of discount rate for the $n=1$ case is unique by the
requirement that it be $\delta_t$.  For $n>1$, the choice is unique
because otherwise the agent's expected discounted utility would not be
$(\rho_t \gamma_t / h) / (1 - \delta_t)$ and thus would not be
independent of $n$.
\eprf

As is standard in economics, for example
in the folk theorem for repeated games~\cite{fudenbergandtiroletext},
we multiply an agent's utility by $(1 - \delta_t)$ so that his
expected utility is independent of his discount rate as well.
With these considerations in mind,
the total expected utility of agent $i$ given the vector of strategies
$\vec{S}$ is
\begin{equation}
\label{eqn:U}
U_i(\vec{S}) = (1 - \delta_{\tau(i)})
\sum_{r = 0}^\infty (1 - (1-\delta_{\tau(i)})/n)^r
E_{\vec{S}}[u_i^r].
\end{equation}

In modeling the game this way, we have implicitly made a number of
assumptions.  For example, we have assumed that all of agent
$i$'s requests that are satisfied give agent $i$ the same utility, and
that prices are fixed.  We discuss the implications of these
assumptions in our companion paper~\cite{scripjournal1}.

Our solution concept is the standard notion of an approximate Nash
equilibrium.
As usual, given a strategy profile $\vec{S}$ and agent $i$, we use
$(S_i',\vec{S}_{-i})$ to denote the strategy profile that is
identical to $\vec{S}$ except that agent $i$ uses $S_i'$.

\dfn \label{def:bestreply}
A strategy $S_i'$ for agent $i$ is an {\em $\epsilon$-best reply}
to a strategy profile $\vec{S}_{-i}$ for the agents other than $i$ in
the game
$(T,\vec{f},h,m,n)$ if, for all strategies
$S_i''$,
$$U_i(S_i'',\vec{S}_{-i}) \leq
U_i(S_i',\vec{S}_{-i}) + \epsilon.$$
\edfn

\dfn \label{def:equilibrium}
A strategy profile $\vec{S}$  for the game\\ $(T,\vec{f},h,m,n)$
is an \emph{$\epsilon$-Nash equilibrium} if for all agents $i$,
$S_i$ is an $\epsilon$-best reply to $\vec{S}_{-i}$.
A \emph{Nash equilibrium} is an epsilon-Nash equilibrium with $\epsilon=0$.
\edfn

As we show in our companion paper~\cite{scripjournal1},
$(T,\vec{f},h,m,n)$ has
equilibria where agents use a particularly simple type of
strategy, called a \emph{threshold strategy}.
Intuitively, an agent with ``too little'' money will want to work, to
minimize the likelihood of running out due to making a long sequence of
requests before being able to earn more money.
On the other hand, a rational agent with plenty of money
will think it is better to delay working, thanks to discounting.
These intuitions suggest that the agent should volunteer if and only if
he has less than a certain amount of money.
Let $s_k$ be the strategy where an agent
volunteers if and only if the requester has at least 1 dollar and the
agent has less than $k$ dollars.
Note that $s_0$ is
the strategy where the agent never volunteers. While everyone
playing $s_0$ is a Nash equilibrium (nobody can do better by
volunteering if no one else is willing to), it is an uninteresting
one.

We frequently consider the situation where each agent of type $t$
uses the same threshold $s_{k_t}$.  In this case, a single vector
$\vec{k}$ suffices to indicate the threshold of each type, and we can
associate with this vector the strategy
$\vec{S}(\vec{k})$ where
$\vec{S}(\vec{k})_i = s_{k_{\tau(i)}}$ (i.e., agent $i$ of type $\tau(i)$
uses threshold $k_{\tau(i)}$).

For the rest of this paper, we focus on threshold strategies (and show
why it is reasonable to do so).
In particular, we show that, if all other agents use threshold
strategies, it is approximately optimal for an agent to use one as
well.  Furthermore there exist Nash equilibria where agents do so.
While there are potentially other equilibria that use different
strategies, if a system designer has agents use threshold strategies by
default, no agent will have an incentive to change.
Since threshold strategies have such low information requirements, they
are a particularly attractive choice for a system designer
as well for the agents, since they are so easy to play.

When we consider the threshold strategy $\vec{S}(\vec{k})$,  for ease
of
exposition, we assume in our analysis that $mhn < \sum_t f_t k_t hn$.
To understand why, note that $mhn$ is the total amount of money in the
system.  If $mhn \ge \sum_t f_t k_t hn$, then if the agents use a
threshold $\vec{S}(\vec{k})$, the system will quickly reach
a state where each agent has $k_t$ dollars, so no agent will
volunteer.  This is equivalent to all agents using a threshold of 0, and
similarly uninteresting.

\section{Summary of Previous Results} \label{sec:summary}

In this section, we summarize the results and definitions from our
companion paper~\cite{scripjournal1} that we use in this paper.
We also provide intuition for the results, some of which is taken from
that paper.
The first theorem shows that there exists a particular distribution
of wealth such that, after a sufficient amount of time, the
distribution of wealth in the system will almost always be close to
that particular distribution.  In order to formalize this statement,
we need a number of definitions.

Let
$$X_{T,\vec{f},h,m,n,\vec{k}} =
\{ \vec{x} \in \mathbb{N}^{hn} \mid \forall i.x_i \leq k_{\tau(i)},
\sum_{i} x_i = hmn \}$$
be the set of allocations of money to agents such that the average
amount of money is $m$ and no agent $i$ has more than $k_{\tau(i)}$
dollars.
The evolution of $\vec{x}^r$ can be described by a Markov chain
$\M_{T,\vec{f},h,m,n,\vec{k}}$ over
the state space $X_{T,\vec{f},h,m,n,\vec{k}}$.
For brevity, we refer to the Markov chain and state space as $\M$
and $X$, respectively, when the subscripts are clear from context.
Let $\Delta_{\vec{f},m,\vec{k}}$ denote the set of probability
distributions $d$ on $\cup_{t \in T} \{t\} \times \prod_t \{ 0,
\ldots, k_t \}$ such that for all types
$t$, $\sum_{l = 0}^{k_t} d(t,l) = f_t$
and $\sum_{t \in T}\sum_{l = 0}^{k_t} l d(t,l) = m$.
We can think of $d(t,l)$ as the fraction of agents
that are of type $t$ and have $l$ dollars.
We can associate each state $\vec{x}$ with its corresponding
distribution $d^{\vec{x}} \in \Delta_{\vec{f},m,\vec{k}}$.
Occasionally, we will make use of distributions $d$ on
$\cup_{t \in T} \{t\} \times \prod_t \{ 0, \ldots, k_t \}$
such that for all types
$t$, $\sum_{l = 0}^{k_t} d(t,l) = f_t$,
without requiring that $\sum_{t \in T}\sum_{l = 0}^{k_t} l d(t,l) = m$;
we denote this set of distributions
$\Delta_{\vec{f},\vec{k}}$.
Given two distributions $d,q \in \Delta_{\vec{f},\vec{k}}$, let
$$H(d||q) = \hspace{.15in}  - \hspace{-.25in} \sum_{\{(t,j) : q(t,j)
\neq 0\}} \hspace{-.25in}  d(t,j) \log
d(t,j)/q(t,j)$$
denote the \emph{relative entropy} of $d$ relative to $q$ ($H(d||q) =
\infty$ if $d(t,j) = 0$ and $q(t,j) \neq 0$ or vice versa); this is
also known as the \emph{Kullback-Leibler divergence of $q$ from
$d$}~\cite{cover}.
For $q$ in $\Delta_{\vec{f},\vec{k}}$, we make use of $d^*_q$, the
distribution in $\Delta_{\vec{f},m,\vec{k}}$ that minimizes relative
entropy relative to $q$.  If $q$ happens to be in
$\Delta_{\vec{f},m,\vec{k}}$ and not just $\Delta_{\vec{f},\vec{k}}$
this is trivially $q$, but it is well defined in general.
Given $\varepsilon > 0$
and $q$, let $X_{T,\vec{f},h,m,n,\vec{k},\varepsilon,q}$
(or $X_{\varepsilon,q}$, for brevity)
denote the set of
states $\vec{x} \in X_{T,\vec{f},h,m,n,\vec{k}}$
such that $\sum_{(t,j)} |d^{\vec{x}}(t,j) - d^*_q|^2 <
\varepsilon$.
Let $I^r_{q,n,\varepsilon}$ be the random variable that is 1 if
$d^{\vec{x}^r} \in X_{\varepsilon,q}$, and 0 otherwise.

\thm \label{thm:distribution}
For all games $(T,\vec{f},h,m,1)$, all vectors $\vec{k}$ of thresholds,
and  all $\varepsilon > 0$,
there exist $q \in \Delta_{\vec{f},\vec{k}}$ and $n_\varepsilon$ such that,
for all $n > n_\varepsilon$,
there exists a round $r^*$  such that, for all $r > r^*$,
we have $\Pr(I^r_{q,n,\varepsilon} = 1) > 1-\varepsilon$. \ethm

Theorem~\ref{thm:distribution} tells us that, after enough time, the
distribution of money is almost always close to some $d^*$, where
$d^*$ can be characterized as a distribution that minimizes relative
entropy subject to
some constraints.
For many of our results, a more explicit characterization will be
helpful.  Let
$q(t,i) = (\omega_t)^i / (\sum_t \sum_{j = 0}^{k_t} (\omega_t)^j)$.
Then the value of $d^*$ is given by the following lemma.

\lem \label{lem:minrelent}
\begin{equation}
\label{eqn:d}
d^*(t,i) = \frac{f_t \lambda^i q(t,i)}
{\sum_{j = 0}^{k_t} \lambda^j q(t,j)},
\end{equation}
where $\lambda$
is the unique value such that
\begin{equation}
\label{eqn:m}
\sum_t \sum_i i d^*(t,i) = m.
\end{equation}
\elem

We now turn from an analysis of the distribution of wealth to an
analysis of best replies and equilibria.
To see that threshold strategies are approximately optimal,
consider a
single agent $i$ of type $t$ and fix the vector $\vec{k}$ of thresholds
used by the other agents.  If we assume that the number of agents is
large, what an agent $i$ does has essentially no affect on the
behavior of the system (although it will, of course, affect that agent's
payoffs).  In particular, this means that the distribution $q$ of
Theorem~\ref{thm:distribution} characterizes the distribution of money
in the system.  This distribution, together with the vector $\vec{k}$ of
thresholds, determines what fraction of agents volunteers at each step.
This, in turn, means that from the perspective of agent $i$, the problem
of finding an optimal response to the strategies of the other agents
reduces to finding an optimal policy in a Markov decision process (MDP)
$\mathcal{P}_{G,\vec{S}(\vec{k}),t}$.  The behavior of the MDP
$\mathcal{P}_{G,\vec{S}(\vec{k}),t}$ depends on
two probabilities: $p_u$ and $p_d$.  Informally,
$p_u$ is the probability of $i$ earning a dollar during each round it
is willing to volunteer, and $p_d$ is the probability that $i$ will be chosen
to make a request during each round.
Note that $p_u$ and $p_d$ depend on aspects of $m$, $\vec{k}$, and $t$;
if the dependence is important, we add the relevant parameters
to the superscript, writing, for example, $p_u^{m,\vec{k}}$.
This MDP is described formally in Appendix~\ref{sec:MDP}.

For many
of our later results and discussion, it will be important to
understand how $p_u$, $p_d$, and $t$ affect the optimal policy for
$\mathcal{P}_{G,\vec{S}(\vec{k}),t}$, and thus the $\varepsilon$-optimal
strategies in the game.  We use this understanding to show that
adding money increases social welfare in Section~\ref{sec:optimize},
to understand how agent behaviors affect social welfare in
Section~\ref{sec:nonstandard}, and to identify agent
types from their behavior in Section~\ref{sec:identify}.  The effects
of these parameters are captured by the following lemma.

\lem \label{lem:br}
Consider the games $G_n = (T,\vec{f},h,m,n)$
(where $T$, $\vec{f}$, $h$, and $m$ are fixed, but $n$ may vary),
and fix the vector $\vec{k}$ of thresholds of the other agents.
There exists a $k$ such that for all $n$, $s_k$ is
an optimal policy for $\mathcal{P}_{G_n,\vec{S}(\vec{k}),t}$.
The threshold
$k$ is the maximum value of $\kappa$ such that
\begin{equation}
\label{eqn:policy}
\alpha_t \leq E[(1 - (1 - \delta_t)/n)^{J(\kappa,p_u,p_d)}] \gamma_t,
\end{equation}

where $J(\kappa,p_u,p_d)$ is a random variable whose value is the first
round in which an agent starting with $\kappa$ dollars,
using strategy $s_\kappa$, and with probabilities $p_u$ and $p_d$ of earning
a dollar and of being chosen given that he volunteers,  respectively,
runs out of money.
\elem

While threshold strategies are optimal for the MDP, given a fixed $p_u$ and
$p_d$, the probabilities $p_u$ and $p_d$ represent the typical long-run
probabilities, not the exact values in each round of the game.
Nevertheless, as
the following theorem shows, threshold strategies are near optimal
in the actual game, not just in the MDP.

\thm \label{thm:threshold}
For all games $G = (T,\vec{f},h,m,n)$, all vectors $\vec{k}$ of
thresholds, and  all $\varepsilon > 0$,
there exist $n^*_\varepsilon$ and $\delta^*_{\varepsilon,n}$
such that for all $n > n^*_\varepsilon$,
types $t \in T$, and $\delta_t > \delta^*_{\varepsilon,n}$,
an optimal threshold policy for
$\mathcal{P}_{G,\vec{S}(\vec{k}),t}$ is an
$\varepsilon$-best reply
to the strategy profile $\vec{S}(\vec{k})_{-i}$
for every agent $i$ of type $t$.
\ethm

Given a game $G = (T,\vec{f},h,m,n)$ and a vector $\vec{k}$ of
thresholds, Lemma~\ref{lem:br} gives an optimal threshold $k_t'$ for
each type $t$.  Theorem~\ref{thm:threshold} guarantees that $s_{k_t'}$
is an $\varepsilon$-best reply to $\vec{S}_{-i}(\vec{k})$, but does
not rule out the possibility of other best replies.  However, for
ease of exposition, we will call $k_t'$ \emph{the} best reply
to $\vec{S}_{-i}$ and call
$\BR_G(\vec{k}) = \vec{k}'$ the best-reply function.  The following
lemma shows that this function is monotone (non-decreasing).
Along the way, several other quantities are shown to be monotone.

\lem \label{lem:monotone}
Consider the family of games\\
$G_m = (T,\vec{f},h,m,n)$ and the
strategies $\vec{S}(\vec{k})$, for $mhn < \sum_t f_t k_t hn$.
For this family of game,
$\lambda_{m,\vec{k}}$ is non-decreasing in $m$ and non-increasing in
$\vec{k}$; $p_u^{m,\vec{k}}$ is non-decreasing in
$m$ and non-increasing in $\vec{k}$; and the function
$\BR_G$ is non-decreasing in $\vec{k}$ and non-increasing in
$m$.
\elem

Our final theorem show that there exists a non-trivial
equilibrium where all agents play threshold strategies
greater than zero.
In the theorem, we refer to the ``greatest'' vector.  By this we
mean that there exists a vector that is an equilibrium and, in a
component-wise comparison, is greater than all other such equilibrium
strategy vectors.  We refer to this particular equilibrium as the
greatest equilibrium.

\thm \label{thm:equilib}
For all games $G = (T,\vec{f},h,m,1)$ and all $\epsilon$, there exist
$n^*_\epsilon$ and $\delta^*_{\epsilon,n}$
such that, if $n > n^*_\epsilon$ and
$\delta_t > \delta^*_{\epsilon,n}$
for all $t$, then there exists a nontrivial vector $\vec{k}$ of thresholds
that is an $\epsilon$-Nash equilibrium.
Moreover, there exists a greatest such vector.
\ethm

The proof of Theorem \ref{thm:equilib} also provides an algorithm for
finding equilibria that seems efficient in practice: start with the
strategy profile $(\infty, \ldots, \infty)$ and iterate the best-reply
dynamics until an equilibrium is reached.

While multiple nontrivial equilibria may exist, in
the rest of this paper, we focus on the greatest equilibrium
in all our applications (although a number of our results hold
for all nontrivial equilibria).  This equilibrium has several
desirable properties, discussed in Section 5 of our companion
paper~\cite{scripjournal1}.

\section{Social Welfare and Scalability}\label{sec:optimize}

In this section, we consider a fundamental question faced by
system designers: what is the optimal amount of money and how does it
depend on the size of the system?  We discuss how  our theoretical
results
summarized in Section~\ref{sec:summary} show that
in order to maximize social welfare, the optimal amount of money is some
constant per agent.  Thus, a system designer that wants to maximize
social welfare should manage the average quantity of money
appropriately.
However, we also show that this must be done carefully.  Specifically,
we show that increasing the amount of money improves performance
up to a certain point, after which the system experiences a monetary
crash.
Once the system crashes, the only equilibrium will be the trivial one
where all agents play $s_0$.  Thus, optimizing the performance of the
system involves discovering how much money the system can
handle before it crashes.

In Section~\ref{sec:model}, we define the game using a
tuple $G = (T,\vec{f},h,m,n)$.  Thus, our definition of a game uses the
average amount of money $m$ rather than the equally reasonable total
amount of money $mhn$.  The choice is motivated by
our theoretical results.  Theorem~\ref{thm:distribution} shows that
the long-term
distribution of money $d^*$ depends on the average amount of money,
but
is independent of $n$,
provided that $n$ is sufficiently large.  Thus, since we
normalize $\delta_t$ by the number of agents in computing utility, the
optimal threshold policy for the MDP developed in
Appendix~\ref{sec:MDP} is also independent of $n$.
Theorems~\ref{thm:threshold}~and~\ref{thm:equilib} show that
such policies constitute an $\varepsilon$-Nash equilibrium.  Thus,
modulo a technical issue regarding the rate of convergence of the
Markov Chain towards its stationary distribution, to determine the
optimal amount of money for a large system, it suffices to determine
the optimal value of $m$, the average amount of money per agent.

We remark that, in practice, it may be easier for the designer to
vary the price of fulfilling a request than to control the
amount of money
in the system.  This produces the same effect. For example, changing
the cost of fulfilling a request from \$1 to \$2 is equivalent to
halving the amount of money that each agent has.  Similarly, halving
the the cost of fulfilling a request is equivalent to doubling the
amount of money that everyone has.  With a fixed amount $hmn$ of money,
there is an optimal product $hnc$ of the number $hn$ of agents and
the cost $c$ of fulfilling a request.

This also tells us how to deal with a
dynamic pool of agents.
Our system can handle newcomers relatively easily: simply allow them
to join with no money.  This gives existing agents no incentive to
leave and rejoin as newcomers.
(By way of contrast, in systems where each new agent starts off with a
small amount of money, such an incentive clearly exists.)
We then change the price of
fulfilling a request so that the optimal ratio is maintained.
This method has the nice feature that it can be implemented in a
distributed fashion; if all nodes in the system have a good estimate
of $n$, then they can all adjust prices automatically.
(Alternatively, the number of agents in the system can be posted in a
public place.)
Approaches that
rely on adjusting the amount of money may require expensive
system-wide computations (see \cite{karma03} for an example), and must be
carefully tuned to avoid creating incentives for agents to manipulate
the system by which this is done.

Note that, in principle, the realization that the cost of fulfilling
a request can change can affect an agent's strategy.  For example,
if an agent expects the
cost to increase, then he may want to defer volunteering to fulfill
a request. However, if the number of agents in the system is always
increasing,
then the cost always decreases, so there is never any advantage in
waiting.
There may be an advantage in delaying a request,
but it is  far more costly, in terms of waiting costs than in
providing service, since we assume the need for a service is often
subject to real waiting costs.
In particular, many service requests, such as those for information or
computation, cannot be delayed without losing most of their value.

Issues of implementation aside, we have now reduced the problem of
determining the optimal
total amount of money for a large system to that of determining the optimal
average amount of money, independent of the exact number of agents.
Before we can determine the optimal value of $m$, we have to
answer a more fundamental question: given an equilibrium that arose
for some value of $m$, how good is it?

Consider a single round of the game with a population
of a single type $t$ and an equilibrium threshold $k$.  If
a request is satisfied,
social welfare increases by $\gamma_t - \alpha_t$; the requester gains
$\gamma_t$ utility and the satisfier pays a cost of $\alpha_t$.
If no request is satisfied then no utility is gained.  What is the
probability that a request will be satisfied?  This requires two
events to occur.  First, the agent chosen to make a request must have
a dollar, which happens with probability approximately $1 - \zeta$,
where $\zeta = d^*(t,0)$ is the fraction of agents with no money.
Second, there
must be a volunteer able and willing to satisfy the request.  Any
agent who does not have his threshold amount of money is willing
to volunteer.  Thus, if $\theta = d^*(t,k_t)$ is the fraction
of agents at their
threshold, then the probability of having a volunteer is
$1 - (1 - \beta_t)^{(1 - \theta)n}$.
We believe that in most large systems this probability is quite close
to 1; otherwise, either $\beta_t$ must be unrealistically small or
$\theta$ must be very close to 1.  For example, even if
$\beta = .01$ (i.e., an agent can satisfy 1\% of requests),
$(1 - \theta)n = 1000$
agents will be able to satisfy 99.99\% of requests.
If $\theta$ is
close to 1, then agents will have an easier time earning money then
spending money (the probability of spending a dollar is at most $1 /
n$, while for large $\beta$ the probability of earning a dollar if an
agent volunteers is roughly $(1 / n)(1 / (1 - \theta))$).  If an agent
is playing $s_4$ and there are $n$ rounds played a day, this means
that for $\theta = .9$ he would be willing to pay $\alpha_t$ today to
receive $\gamma_t$ over 10 years from now.
For most systems, it seems
unreasonable to have $\delta_t$ or $\gamma_t / \alpha_t$ this large.
Thus, for the purposes of our analysis, we
approximate $1 - (1 - \beta_t)^{(1 - \theta)n}$ by 1.

With this approximation, we can
write the expected increase in social welfare each round as
$(1 - \zeta)(\gamma_t - \alpha_t)$.
If we have more than one type of agent, the situation is essentially the
same.  The
equation for social welfare is more complicated because now the gain in
welfare depends on the $\gamma$, $\alpha$, and $\delta$ of the agents
chosen in that round, but the overall analysis is the same, albeit with
more cases.
In the general case,
\begin{equation}
\label{eqn:zeta}
\zeta = \sum_t d^*(t,0)
\end{equation}
Thus our goal is clear: find the amount of money that,
in equilibrium, minimizes $\zeta$.

In general, as the following theorem shows, $\zeta$ decreases as $m$
increases.  More specifically, given our assumption that the system is
starting at the greatest equilibrium $\vec{k}$, increasing $m$ and
then following best response dynamics leads to the new greatest
equilibrium $\vec{k}'$.  As long as $\vec{k}'$ is non-trivial,
$\zeta_{m',\vec{k}'} \leq \zeta_{m,\vec{k}}$.

Theorems~\ref{thm:distribution},~\ref{thm:threshold},~and~\ref{thm:equilib}
place requirements on the values of $n$ and $\delta_t$.
Intuitively, the theorems require
that the $\delta_t$s is sufficiently large to ensure that
agents are patient enough that their decisions are dominated by
long-run behavior rather than short-term utility,
and that $n$ is sufficiently large to
ensure that small changes in the distribution of money do not move it
far from $d^*$.  In the theorems in this section,
assume that these conditions are satisfied.
To simplify the statements of the theorems,
we use ``the standard conditions hold'' to mean that the game
$G = (T,\vec{f},h,m,n)$
under consideration is such that $n > n^*$
and $\delta_t > \delta^*$ for the $n^*$ and $\delta^*$
needed for the results of
Theorems~\ref{thm:distribution},~\ref{thm:threshold},~and~\ref{thm:equilib}
to apply.

\thm \label{thm:money}
Let $G = (T,\vec{f},h,m,n)$ be such that the standard
conditions hold, and let $\vec{k}$ be
the greatest equilibrium for $G$.
Then if $m' > m$, the best-reply dynamics in $G' = (T,\vec{f},h,m',n)$
starting at $\vec{k}$ converge to some
$\vec{k}' \leq \vec{k}$ that is the greatest equilibrium of $G'$.
If $\vec{k}'$ is a nontrivial equilibrium,
then $\zeta_{m',\vec{k}'} \leq \zeta_{m,\vec{k}}$.
\ethm

\prf
In the proof of Theorem~\ref{thm:equilib}, it is shown that starting
at any vector $\vec{k_0}$ greater than the greatest equilibrium and
applying best-reply dynamics (iteratively replacing $\vec{k_i}$ with the
vector of best-reply strategies $\vec{k_{i+1}} = \BR_G(\vec{k_i})$)
leads to the greatest equilibrium in a finite number of steps.
Since $\vec{k}$ is an equilibrium, $\BR_G(\vec{k}) = \vec{k}$.
By Lemma~\ref{lem:monotone}, $\BR_G$ is non-increasing in $m$.
Thus, $\vec{k} = \BR_G(\vec{k}) \geq \BR_{G'}(\vec{k})$.
Applying best-reply dynamics using $\BR_{G'}$ starting at
$\vec{k}$
gives us an equilibrium
$\vec{k}'$ such that $\vec{k}' \le \vec{k}$.
By Lemma~\ref{lem:monotone}, $\BR_G(\vec{k}'')$
is non-decreasing in $\vec{k}''$,
so this is the greatest equilibrium.
Suppose that $\vec{k}'$ is nontrivial.
By Equations~(\ref{eqn:d})~and~(\ref{eqn:zeta}),
$$\zeta_{m,\vec{k}} = \sum_t d^*(t,0) = \sum_t
\frac{f_t \lambda_{m,\vec{k}}^i q(t,i)}
{\sum_{j=0}^{k_t} \lambda_{m,\vec{k}}^j q(t,j)}.$$
Again by Lemma~\ref{lem:monotone}, $\lambda_{m,\vec{k}}$ is
non-decreasing in $m$ and non-increasing in $\vec{k}$.  Thus,
$\zeta_{m',\vec{k}'} \leq \zeta_{m,\vec{k}}$.
\eprf

\commentout{
Theorem \ref{thm:money} makes several strong statements about what
happens to social welfare as the amount of money increases (assuming
there is no monetary crash).  Taking the worst-case view, we know
social welfare at the maximum equilibrium will increase.  It also says
that if the system is jolted out of the greatest
equilibrium by a sudden addition of money and agents
use best-reply dynamics and find a new nontrivial equilibrium,
social welfare will have increased.  This gives the system designer
confidence that the system will not end up in a bad
equilibrium if she adjusts the amount of money so as to increase the
average amount of money.	
}

Theorem~\ref{thm:money} tells us that, as long as the system does not
crash, more money is better.  The following corollary tells us that
such a crash is an essential feature;
a sufficient increase in the amount of
money leads to a monetary crash.  Moreover, once the system has
crashed, adding more money does not cause the system to become
``uncrashed.''

\cor \label{cor:crash}
Consider the family of games\\
$G_m = (T,\vec{f},h,m,n)$ such that the
standard conditions hold.
There exists a critical average amount $m^*$ of money  such that
if $m < m^*$, then $G_m$ has a nontrivial equilibrium, while if $m >
m^*$, then $G_m$ has no nontrivial equilibria
in threshold strategies.
(A nontrivial equilibrium may or may not exist if $m=m^*$.)
\ecor
\prf
To see that there is some $m$ for which $G_m$ has no nontrivial
equilibrium, fix $m$.  If there is no nontrivial
equilibrium in $G_m$, we are done.  Otherwise, suppose that the greatest
equilibrium in $G_m$ is $\vec{k}$.
Choose $m' > \sum_t f_t k_t$, and let $\vec{k}'$ be the
greatest equilibrium in $G_{m'}$.
By Theorem~\ref{thm:money}, $\vec{k}' \leq \vec{k}$.
But if $\vec{k}'$ is a nontrivial equilibrium then, in
equilibrium, each agent of type $t$ has at most
$k'_t \leq k_t$ dollars.
But then $m' >  \sum_t f_t k_t \ge \sum f_t k_t'$, a contradiction.

Let $m^*$ be the infimum over all $m$ for
which no nontrivial equilibrium exists in the game $G_m$.
(We know that $m^* > 0$ by Theorem~\ref{thm:equilib}.)
Clearly, by choice of $m^*$, if $m < m^*$, there is
a nontrivial equilibrium in $G_m$.  Now suppose that
$m > m^*$.  By the construction of
$m^*$, there exists $m'$ with $m > m' \geq m^*$ such
that no
nontrivial equilibrium exists in $G_{m'}$.
Let the
greatest equilibria with $m'$ and $m$ be  $\vec{k}'$ and
$\vec{k}$, respectively.
By Theorem~\ref{thm:money}, $\vec{k} \leq \vec{k}'$.
Thus $\vec{k}$ is also trivial.
\eprf

\begin{figure}[htb]
\centering \epsfig{file=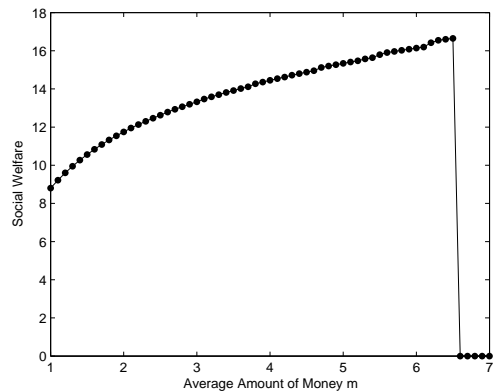, height=2.2in}
\caption{Social welfare for various average amounts of money,
demonstrating a monetary crash.}
\label{fig:crash}
\end{figure}

Figure \ref{fig:crash} shows an example of the monetary crash in the
game
with two types of agents (with parameters
$\alpha_t,\beta_t,\gamma_t,\delta_t,\rho_t,\chi_t$) where thirty
percent are of the first type, there are 10 agents in the base
economy, 100 replicas, and the average amount of money is $m$.
Formally, the game is:
$$(\{(.05,1,1,.95,1,1),(.15,1,1,.95,1,1) \},(.3,.7),10,m,100).$$

\begin{figure}[htb]
\centering \epsfig{file=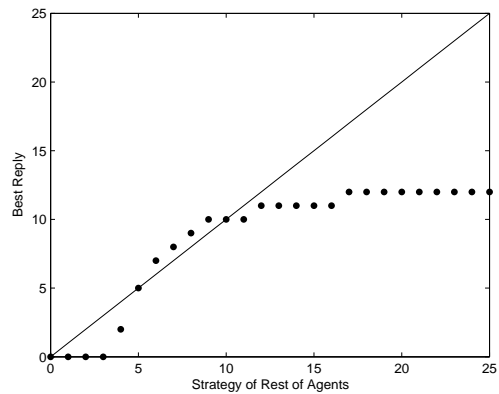, height=2.2in}
\caption{A hypothetical best-reply function with one type of agent.}
\label{fig:br}
\end{figure}

Corollary~\ref{cor:crash} tells us that this crash is a very sharp
phenomenon; with some amount of money the system performs well, but
with just slightly more the system stops working.  This is a result of
the way increasing the amount of money affects agent's best reply
functions.
Figure~\ref{fig:br}, reproduced from our companion
paper~\cite{scripjournal1}, gives an example of a best reply
function with one type of agent.
It shows, for a particular fixed value of $m$, how the
optimal strategy for an agent depends on the the strategies of the
other agents.  Thus, an equilibrium is a point on the line $y=x$: the
optimal strategy is exactly the strategy the other agents are using.
In this simple case, increasing $m$ causes every point to shift
downward (since strategies are discrete, there may be some minimum
increase for a particular point to shift).  With a large enough
increase in $m$, every point except (0,0) will be below the line and
so there will be no nontrivial equilibrium.  The sharpness is a result
of there being a critical value $m^*$ at which the last point drops
below the line.

In light of Corollary~\ref{cor:crash}, the system
designer should try to find $m^*$, the point where social welfare is
maximized.
In doing so, it is helpful to have an understanding of the types and
strategies of agents in the systems, an issue we discuss in
Section~\ref{sec:identify}.
In practice, the system designer may want to set the average amount
of money $m$ to be somewhat less than
$m^*$.  Since there will be a crash if  $m > m^*$, small
changes in the characteristics of the population or mistakes by the
designer in modeling them could lead to a crash if she chooses $m$
too close to $m^*$.

The phenomenon of a monetary crash is intimately tied to our
assumption of fixed prices.  We saw such a crash in practice in the
babysitting co-op example.  If the price is allowed to float freely,
we expect that, as the money supply increases,
there will be inflation; the price will increase so as to avoid a
crash.  However, floating prices can create other monetary problems,
such as speculation, booms, and busts.
Floating prices also impose transaction costs on agents.  In systems
where prices would normally be relatively stable, these transaction
costs may well outweigh the benefits of floating prices, so a system
designer may opt for fixed prices, despite the
risk of a crash.

We believe there may also be a happy medium between a single, permanent
fixed price and prices that change freely from round to round; indeed,
our advice to system designers points naturally toward it.
In particular,
our advice about how to optimize the amount of money relies on
experimentation and observation to determine what agents are doing and
what their utilities are.  This information then tells the designer
how much money she should provide.  Since adjusting the amount of
money is equivalent to adjusting prices, the designer could
incorporate this process into a price setting rule.  Depending on the
nature of the system, this could either be done manually over time (if
the information is difficult to gather and analyze) or
automatically (if the information gathering and analysis can itself be
automated).  From this perspective, a monetary crash, though real, is
not something to be feared.  Instead, it is just a strong signal that
the current price, while probably not too far off from a very good
price, requires adjustment.  Naturally, this relies on a process that
proceeds slow enough that agents myopically ignore the effects of
future price changes in determining their current action.

Finally, a system
designer could consider interventions other than adjusting the amount
of money.  One obvious opportunity is the process by which volunteers
are selected.  Our model assumes this process is random, but it need
not be.  For example, the system designer could attempt to bias the
process in favor of agents with smaller amounts of money.  Like increasing
the average amount of money, this could increase efficiency since
agents would spend less time with no money, but could potentially
cause a crash, since agents have less of an incentive to save for the
future.  Our techniques rely on the choice of agents being independent
of how much money they have,  do not allow us to rigorously analyze this
situation.  Biasing the volunteer
selection rule is an idea we return to in Section~\ref{sec:sybil}, as
agents who create sybils increase their probability of being chosen.

\section{Dealing with Nonstandard Agents} \label{sec:nonstandard}

The model in Section~\ref{sec:model} defines the utility of
standard agents, who value service and dislike using their resources to
provide it to
others.  This seems like a natural description of the way most people
use distributed systems.  However, in a real system, not every user
will behave they way the designer intends.  A practical system needs
to be robust to nonstandard behaviors.  In this section, we show how
our model can be used to understand the effects of four interesting
types of nonstandard behavior.  First, an agent might provide service
even when he will receive nothing in return, behaving as an altruist.
Second, rather than viewing money as a means to satisfy future
requests, an agent might place an inherent value on it and start
hoarding it.  Third, an agent might create additional identities,
known as \emph{sybils}, to try and manipulate the system. Finally, agents
might collude with each other.

The results of this section give a system designer insight into how to
design a scrip system that takes into account (and is robust to) a number of
frequently-observed behaviors.

\subsection{Altruists} \label{sec:altruists}

P2P filesharing systems
often have large numbers of free riders; they work because a small
number of altruistic users provide most of the files.
For example, Adar and Huberman~\citeyear{adar00} found that, in the
Gnutella network, nearly 50 percent of responses are from the top 1
percent of sharing hosts.  A wide variety of systems have been
proposed to discourage free riding.
However, according to our model, unless this system mostly
eliminates the altruistic users, adding such a system
will have no effect on rational users.

To make this precise, take an altruist to be someone who always volunteers
to fulfill requests, regardless of whether the other agent can pay.
Agent $i$ might rationally behave altruistically if,
rather than suffering a loss of utility when satisfying a
request, $i$ derives positive utility from satisfying it.
Such a utility function is a reasonable representation of the
pleasure that some people get from the sense that they provide
the music that everyone is playing.  For such altruistic agents,
the strategy of always volunteering is dominant.
While having a nonstandard utility function might be one reason that
a rational agent might use this strategy,
there are certainly others.  For example a naive user of filesharing
software with a good connection might well follow this
strategy.
All that matters for the
following discussion is that there are some agents that use this
strategy, for whatever reason.  For simplicity, we assume that all
such agents have the same type $t_a$.

Suppose that a system has $a$ altruists.
Intuitively, if $a$ is moderately large, they
will manage to satisfy most of the requests in the system even if
other agents do no work. Thus, there is little incentive for any
other agent to volunteer, because he is already getting full
advantage of participating in the system. Based on this intuition,
it is a relatively straightforward calculation to determine
a value of $a$
that depends only on the types, but not the number
$n$, of agents in the system,
such that the dominant strategy for all standard
agents $i$ is to never volunteer to satisfy any requests.

\pro \label{pro:altruist}
For all games $(T,\vec{f},h,m,1)$
with $f_{t_a} > 0$, there exists a value $a$ such that, if $n > a / (f_{t_a}
h)$
(i.e., there are at least $a$ altruists),
then never volunteering is a dominant strategy
for all standard agents. \epro

\prf
Consider the strategy for a standard
agent $i$ in the presence of $a$ altruists.
Even with no money, agent $i$ will get a request
satisfied with probability $1 - (1 - \beta_{t_a})^a$
just through the actions of the altruists.
Consider a round when agent $i$ is chosen to make a request.
If he has no money (because he never volunteered), his expected utility
is $\gamma_{\tau(i)} (1 - (1 - \beta_{t_a})^a)$.
His maximum possible utility for the round is $\gamma_{\tau(i)}$.  Thus, a
strategy where he volunteers can increase his utility for a round by at
most $\gamma_{\tau(i)} (1 - \beta_{t_a})^a$.
Thus, even if the agent gets every request satisfied,
his expected utility can increase by at most
$$\begin{array}{ll}
&(1 - \delta_{\tau(i)}) \sum_{r=0}^\infty \frac{\rho_{\tau(i)}}{hn}
\gamma_{\tau(i)} (1-\beta_{t_a})^a (1-\frac{1-\delta_{\tau(i)}}{n})^r\\
= & (1 - \delta_{\tau(i)}) (\rho_{\tau(i)} / h)
\gamma_{\tau(i)} (1 - \beta_{t_a})^a / (1 - \delta_{\tau(i)})\\
= & (\rho_{\tau(i)} / h) \gamma_{\tau(i)} (1 - \beta_{t_a})^a.
\end{array}$$
Clearly this expression goes to 0 as $a$ goes to infinity.  If we take
$a$ large enough that the expression is less than $\alpha_t$ for all
types $t$, then the value of having every future request satisfied is
less than the cost of volunteering now, so no agent will ever volunteer.
\eprf

Consider the following reasonable values for our parameters: $\beta_t
= .01$ (so that each player can satisfy 1\% of the requests),
$\gamma_t = 1$,
$\alpha_t = .1$ (a low but non-negligible cost), $\delta_t = .9999$/day
(which
corresponds to a yearly discount factor of approximately $0.95$),
and an average of 1 request per day per player.
Then as long as $a > 1145$ to ensure that not volunteering is a dominant
strategy.  While this is a large number, it is small
relative to the
size of a large P2P network.
While the number of altruists needed to degrade the performance of the
system increases somewhat with the number of agents,  the point remains
that a small
fraction of altruists can discourage the rest of the system from
providing service.

Proposition~\ref{pro:altruist} shows that with enough altruists, the
system eventually experiences a monetary crash, since all agents will use
a threshold of zero.  However, interesting behavior can still arise
with smaller numbers of altruists.
consider the situation where an $a$ fraction of requests are
immediately satisfied at no cost without the requester needing to ask
for volunteers.
Intuitively, these are the
requests satisfied by the altruists, although
the following result also applies to
any setting where agents occasionally have a (free) outside option.
The following theorem shows that social welfare is increasing in $a$.

Let $G = (\{t\},1,h,m,n)$ be a game with a single type for which the
standard conditions hold.   Consider the family $G_a$ of games
(parameterized by $a$) that result from
$G$ if a fraction $a$ of requests can be satisfied at no cost.  That is,
the game $G_a$ is the same as $G$, except that if an agent $i$ makes a
request, with probability $a$, it is satisfied at no cost, and with
probability $1-a$, an agent is chosen among the volunteers to satisfy
the request, just as in $G$, and the $i$ is charged 1 dollar to have the
request satisfied.

\thm \label{thm:altruists}
For the interval of values of $a$ where there is no monetary crash in
$G_a$,
social welfare increases as $a$ increases (assuming
that the greatest equilibrium is played by all agents in $G_a$).
\ethm

\prf
An agent's utility in a round where he makes a request and it is
satisfied at no cost is $\gamma_t$.  Since such rounds occur with
probability $a$, by assumption, our normalization guarantees that
the sum of standard agents' expected utilities in rounds
where a request is satisfied at no cost is $a \gamma_t$.
The same analysis as in Section \ref{sec:optimize} shows that the
sum of agents' expected utilities in each of the remaining rounds is
$(1-a)(1 - \zeta(a))(\gamma_t - \alpha_t)$,
where, as before, $\zeta(a) = d^*(t,0,a)$, the equilibrium value of
$d^*(t,0)$ in the game $G_a$.
Thus, expected utility as a function of $a$ is
\begin{equation}
\label{eqn:alt}
a \gamma_t + (1-a)(1 - \zeta(a))(\gamma_t - \alpha_t).
\end{equation}
To see that this expression increases as $a$ increases, we would like to
take the derivative relative to $a$ and show it is positive.
Unfortunately, $\zeta(a)$ may not even be continuous.  Because strategies
are integers, there will be regions where $\zeta(a)$ is constant, and
then a jump when a critical value of $a$ is reached that causes the
equilibrium to change.
At a point $a$ in a region where $\zeta(a)$ is constant,
$\zeta'(a) = 0$, so the derivative of
Equation~(\ref{eqn:alt}) is
$\gamma_t - (1 - \zeta(a))(\gamma_t - \alpha_t) > 0$.  Hence,  social welfare
is increasing at
such points.

Now consider a point $a$ where $\zeta(a)$ is discontinuous.
Such a discontinutity occurs when the
greatest equilibrium, the greatest value $\vec{k}$ for which
$\BR_{G_a}(\vec{k}) = \vec{k}$, changes.
We show that, for a fixed $\vec{k}$,
$\BR_{G_a}(\vec{k})$ is non-increasing in $a$.
Since increasing $a$ can only cause the $\BR_{G_a}(\vec{k})$ to
decrease, the discontinuity must be caused by a change from an
equilibrium $\vec{k}$ to a new equilibrium $\vec{k}' < \vec{k}$.
Fix a vector $\vec{k}$ of thresholds, and let $p_u^{\vec{k},m,a}$ be the
probability that $i$ will earn a dollar in a given round if he is
willing to volunteer, given that a fraction $a$ of requests is satisfied
at no cost (so that $p_u^{\vec{k},m,0}$ is what we earlier called
$p_u^{\vec{k},m}$); we similarly define $p_d^{\vec{k},m,a}$,
his probability of being chosen to make a request.
It is easy to see that $p_u^{\vec{k},m,a} = (1 - a) p_u^{\vec{k},m,0}$
and $p_d^{\vec{k},m,a} = (1 - a) p_d^{\vec{k},m,0}$.
The random variable $J(\kappa,p_u,p_d)$
in Equation~(\ref{eqn:policy}) describes
the first time at which an agent starting with $\kappa$ dollars and using
the threshold $\kappa$ while earning a dollar with probability $p_u$ and
spending a dollar with probability $p_d$ reaches zero dollars.  As $a$
increases, $p_u^{\vec{k},m,a}$ and $p_d^{\vec{k},m,a}$ both decrease,
but the ratio $p_u^{\vec{k},m,a}/p_d^{\vec{k},m,a}$ remains constant.
Intuitively, this means that the agent ``slows down'' his random walk
on amounts of money by a factor of $1 / (1-a)$.
This slowdown occurs because, each time the agent would have an
opportunity to volunteer or would have spent a dollar, with
probability $a$ the opportunity is taken by an altruist instead, so
the expected time to take a step increases by a factor of $1/(1-a)$.
Thus, the value of the expectation in Equation~(\ref{eqn:policy}),
and hence the right-hand side of Equation~(\ref{eqn:policy}),
decreasing as a function of $a$.
By Lemma~\ref{lem:br}, $(\BR_{G_a}(\vec{k}))_t$ is the maximum value
of $\kappa$ such that Equation~(\ref{eqn:policy}) is satisfied.
Decreasing the right-hand side can only decrease the maximum value of
$\kappa$, so
$\BR_{G_a}(\vec{k})$ is
non-increasing
as a function of $a$.

By Lemma~\ref{lem:monotone},
$\lambda_{m,\vec{k}}$ is non-increasing in $\vec{k}$
(unless the system crashes,
after which it remains crashed even when a is increased further, as in
Corollary~\ref{cor:crash}, where the points at which social welfare is
increasing form an interval).
Since, as we have just shown, if there is
a discontinuity at $\zeta(a)$ when $a$ increases, the greatest
equilibrium changes
at $a$ from $\vec{k}$ to $\vec{k}' < \vec{k}$,
we must have
$\lambda_{m,\vec{k}'} \geq \lambda_{m,\vec{k}}$.
In Equation~(\ref{eqn:d}) for $i = 0$, the value of the numerator is
independent of $\lambda$, but the denominator with
$\lambda_{m,\vec{k}'}$ is greater than or equal to the denominator with
$\lambda_{m,\vec{k}}$.  Thus $d^*(t,0,a) = \zeta(a)$ is
non-increasing at $a$.
By Equation~(\ref{eqn:alt}), this means that expected utility is
increasing at $a$. Thus, in either case, social welfare is
increasing in $a$.
\eprf

\begin{figure}
\centering
\includegraphics[height=2.2in]{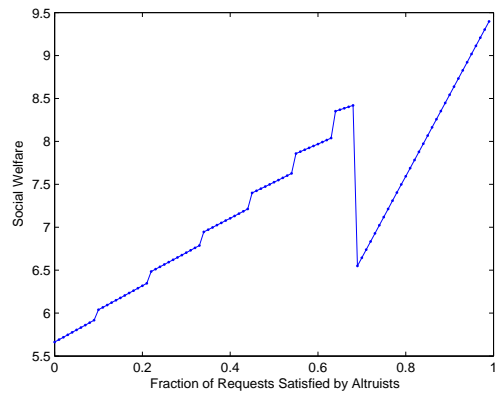}
\caption{Altruists can cause a crash.}
\label{fig:altruism}
\end{figure}

\begin{figure}
\centering
\includegraphics[height=2.2in]{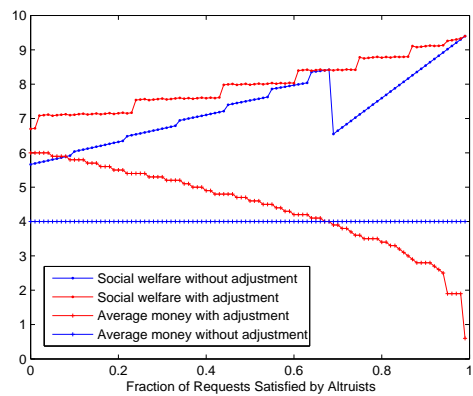}
\caption{$m$ can be adjusted as $a$ increases.}
\label{fig:altruism_adjusted}
\end{figure}

Theorem~\ref{thm:altruists} and Proposition~\ref{pro:altruist}
combine to tell us that a little altruism is good
for the system, but too much causes a crash.
Figure~\ref{fig:altruism} demonstrates this phenomenon.  As we saw in
Section~\ref{sec:optimize}, such crashes are caused when $m$, the
average amount of money,  is too
large.  By decreasing $m$ appropriately, even relatively large values
of $a$ can be exploited, as Figure~\ref{fig:altruism_adjusted} shows.
The ``social welfare without adjustment'' plot is the same data from
Figure~\ref{fig:altruism}, with the corresponding plot of the amount
of money horizontal since $m$ was held fixed.
By decreasing the average amount of money appropriately as the number of
altruists increases, a system designer can increase social welfare while
avoiding a crash
of the economy (the system will still function due to the presence of
altruists).
Note that, in discussing social welfare, our formulation excludes the
welfare of the altruists, since our focus here is on the effects of
altruism on standard agents.


\subsection{Hoarders}

Whenever a system allows agents to accumulate something, be it work
done, as in SETI@home, friends on online social networking sites, or
``gold'' in an online game, a certain group of users seems to make it
their goal to accumulate as much of it as possible.  In pursuit of
this, they will engage in behavior that seems
irrational.
For simplicity here, we model hoarders as playing the strategy
$s_{\infty}$.  This means that they will volunteer under all
circumstances.  Our analysis would not change significantly if we also
required that they never made a request for work.  Our first result
shows that, for a fixed money supply, having more hoarders makes
standard agents worse off.

Consider a game $G = (T,\vec{f},h,m,n)$ such that the standard
conditions hold. Consider the family $G_{f_h}$ of games
(parameterized by $f_h$) that result from $G$ if
a fraction $f_h$ of agents are
hoarders.  That is, $G_{f_h} = (T \times \{0,1\},\vec{f}',h',m,n)$
where an agent of type $(t,0)$ is a standard agent of type $t$, but an
agent of type $(t,1)$ is a hoarder and always uses the strategy
$s_\infty$ (his probabilities are still determined by $\beta_t$,
$\rho_t$, and $\chi_t$).  Define $\vec{f}'$ by taking
$f_{(t,0)}' = (1 - f_h)f_t$ and $f_{(t,1)}' = f_hf_t$ for all types $t$.
Let $h'$ be the smallest
multiple of $h$ such that $f_{(t,i)}h'$ is an integer for all
$t$ and $i$.  (We need to adjust $h$ because otherwise the number of
agents in the base game may not be well defined.)  Finally, to account
for the changed $h$, let
$\delta_{(t,i)} = 1 - (1 - \delta_t)h/h'$.

\thm\label{thm:hoarders}
In the family $G_{f_h}$ of games,
social welfare is non-increasing in $f_h$ (if the greatest
equilibrium is played by all agents in $G_{f_h}$).
\ethm

\prf
Let $\vec{k}(f_h)$ denote the greatest equilibrium in $G_{f_h}$.
An increase in $f_h$ is equivalent to taking some number of standard
agents and increasing their strategy to $s_\infty$.
It follows from Lemma~\ref{lem:monotone} that $\BR_{G_{f_h}}$ is
non-decreasing  in $f_h$, and so $\vec{k}(f_h)$ is non-decreasing
in $f_h$.
Again by Lemma~\ref{lem:monotone},
$\lambda_{m,\vec{k}(f_h)}$ is non-increasing in $f_h$.
Let $\zeta{f_h} = 1 / (1 - f_h) \sum_t d^*((t,0),0,f_h)$ be the
fraction of non-hoarders with zero dollars, where $d^*((t,0),0,f_h)$
is the value of $d^*((t,0),0)$ at the greatest equilibrium of $G_{f_h}$.
By Equation~(\ref{eqn:d}), $\zeta(f_h)$ is non-decreasing in $f_h$.
Thus, social welfare is non-increasing in $f_h$.
\eprf

Hoarders do have a beneficial aspect.  As we have observed, a monetary
crash occurs when a dollar becomes valueless, because there
are no agents willing to take it.  However, with hoarders in the system,
there is always someone who will volunteer, so there cannot be a
crash.  Thus, for any $m$, the greatest equilibrium will be nontrivial
and, by Theorem~\ref{thm:money}, social welfare keeps increasing as $m$
increases.
So, in contrast to altruism, where the appropriate response was to
decrease $m$, the appropriate response to hoarders is to increase
$m$.  In fact, our results indicate that the optimal response to
hoarders is to make $m$ infinite.  This is due to our unrealistic
assumption that hoarders would use the strategy $s_\infty$ regardless
of the value of $m$.  There is likely an upper limit on the value
of $m$ in practice, since it is unlikely that hoarders would be willing
to hoard scrip if it is so easily available.

\subsection{Sybils} \label{sec:sybil}

Unless identities in a system are tied to a real world identity (for
example by a credit card), it is effectively impossible to prevent a
single agent from having multiple identities~\cite{sybil}.
Nevertheless, there are a number of techniques that can make it
relatively costly for an agent to do so.  For example, Credence uses
cryptographic puzzles
to impose a cost each time a new identity wishes to
join the system \cite{credence}.  Given that a designer can impose
moderate costs to sybilling, how much more need she worry about the
problem?  In this section, we show that the gains from creating sybils
when others do not diminish rapidly, so modest costs
may well be sufficient to deter sybilling by typical users.
However, sybilling is a self-reinforcing phenomenon.  As the number
of agents with sybils gets larger, the cost to being a non-sybilling
agent increases, so the incentive to create sybils becomes stronger.
Therefore, measures to discourage or prevent sybilling should be taken
early before this reinforcing trend can start.  Finally, we examine the
behavior of systems where only a small fraction of agents have sybils.
We show that under these circumstances a wide variety of outcomes are
possible (even when all agents are of a single type), ranging from a
crash (where no service is provided) to an
increase in social welfare.
This analysis provides insight into the tradeoffs between
efficiency and stability that occur when controlling the money supply
of the system's economy.

When an agent of type $t$ creates sybils, the only parameter of his type
that may change as a result is $\chi_t$, if we redefine the likelihood of an
agent being chosen to be the likelihood of the agent or any of his
sybils being chosen.
The other parameters, such as $\rho$, remain unchanged because there
is no particular reason that having multiple identities should cause
the agent to, for example, desire service more often.
For simplicity, we assume that each sybil is
as likely to be chosen as the original agent, so creating $s$ sybils
increases $\chi_t$ by $s \chi_t$.
(Sybils may have other impacts on the system, such as increased search
costs, but we expect these to be minor.)

\begin{figure}
\centering
\includegraphics[height=2.2in]{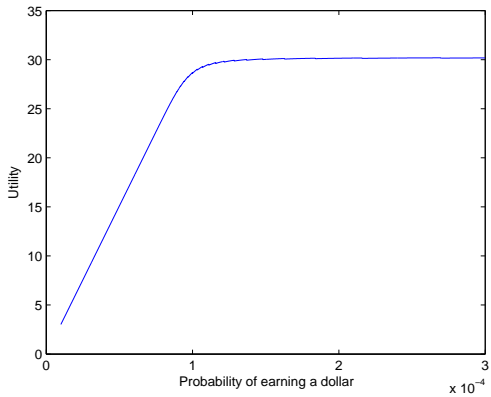}
\caption{The effect of $p_u$ on utility}
\label{fig:sybilben}
\end{figure}

Increasing $\chi_t$ benefits an agent by increasing his value of
$\omega_t$ and thus $p_u$, his probability of earning a dollar
(see Equation~(\ref{eqn:pu}) in Appendix~\ref{sec:MDP}).
When $p_u < p_d$, the agent has more opportunities to
spend money than to earn money, so he will regularly have requests go
unsatisfied due to a lack of money.
In this case, the fraction of
requests he has satisfied is roughly $p_u / p_d$, so increasing
$p_u$ by creating sybils
results in a roughly linear increase in utility.
As Theorem~\ref{thm:peps} shows, when $p_u$ is
close to $p_d$, the increase in satisfied requests is no longer
linear, so the benefit of increasing $p_u$ begins to diminish.
Finally, when $p_u > p_d$, most of the agent's requests are being
satisfied, so the benefit from increasing $p_u$ is very small.
Figure~\ref{fig:sybilben} illustrates an agent's utility as $p_u$
varies for $p_d = .0001$.%
\footnote{
Except where otherwise noted, the remaining figures in this section
assume that $m = 4$, $n = 10000$ and that there is
a single type of rational agent with $\alpha = .08$, $\beta = .01$,
$\gamma = 1$, $\delta = .97$, $\rho = 1$, and $\chi = 1$.
These values
are chosen solely for illustration, and are
representative of a broad range of parameter values.
The figures are based on calculations of the equilibrium behavior.}
We formalize the relationship between $p_u$, $p_d$, and the agent's
utility in the following theorem, whose proof is deferred to
Appendix~\ref{sec:nonstandard_proofs}.

\begin{figure}
\centering
\includegraphics[height=2.2in]{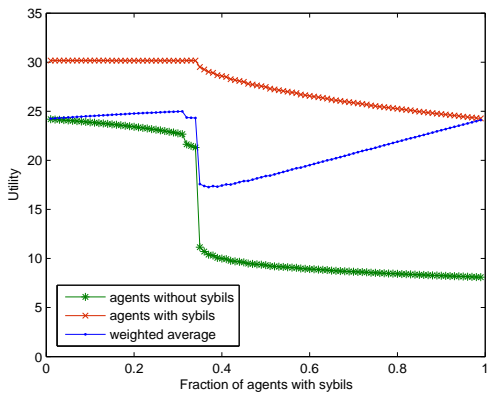}
\caption{The effect of sybils on utility}
\label{fig:sybilcost}
\end{figure}

\thm \label{thm:peps}
Fix a game $G$ and vector of thresholds $\vec{k}$.
Let $R_{\vec{k},t} = p_u^{\vec{k},t} / p_d^t$.
In the limit as the number of rounds goes to
infinity, the fraction of the agent's requests that have an agent
willing and able to satisfy them that get satisfied is
$(R_{\vec{k},t} - R_{\vec{k},t}^{k_t+1}) / (1 - R_{\vec{k},t}^{k_t+1})$
if $R_{\vec{k},t} \neq 1$ and
$k_t / (k_t + 1)$ if $R_{\vec{k},t} = 1$.
\ethm

Theorem~\ref{thm:peps} gives insight into the
equilibrium behavior with sybils.  Clearly, if sybils have no
cost, then creating as many as possible is a dominant strategy.
However, in practice, we expect there is some modest overhead involved
in creating and maintaining a sybil, and that a designer can take steps
to increase this cost without unduly burdening agents.  With such a
cost, adding a sybil might be valuable if $p_u$ is much less than
$p_d$, and a net loss otherwise.  This makes sybils a
self-reinforcing phenomenon.  When a large number of agents create
sybils, agents with no sybils have their $p_u$ significantly
decreased.  This makes
them much worse off and makes sybils much more attractive to them.
Figure~\ref{fig:sybilcost} shows an example of this effect.
This self-reinforcing quality means that it is important to take steps to
discourage the use of sybils before they become a problem.  Luckily,
Theorem~\ref{thm:peps} also suggests that a modest cost to create
sybils will often be enough to prevent agents from creating them
because with a well chosen value of $m$, few agents should have low
values of $p_u$.

We have interpreted Figures~\ref{fig:sybilben}~and~\ref{fig:sybilcost}
as being about changes in $\chi$ due to sybils, but the results hold
regardless of what caused differences in $\chi$.  For example, agents
may choose a volunteer based on characteristics such as connection
speed or latency.
If these characteristics are
difficult to verify and do impact decisions, our results show that
agents have a strong incentive to lie about them.
This also suggests that the decision about what sort of information
the system should enable agents to share involves tradeoffs.  If
advertising legitimately allows agents to find better service or more
services they may be interested in, then advertising can increase social
welfare.  But if
these characteristics impact decisions but have little impact on the
actual service, then allowing agents to advertise them can lead to a
situation like that in Figure~\ref{fig:sybilcost}, where some agents
have a significantly worse experience.

\begin{figure}
\centering
\includegraphics[height=2.2in]{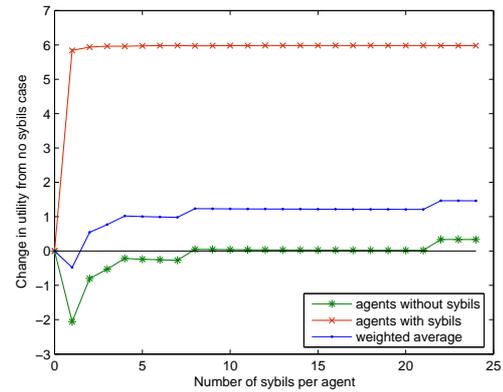}
\caption{Sybils can improve utility}
\label{fig:sybilsmall}
\end{figure}
\begin{figure}
\centering
\includegraphics[height=2.2in]{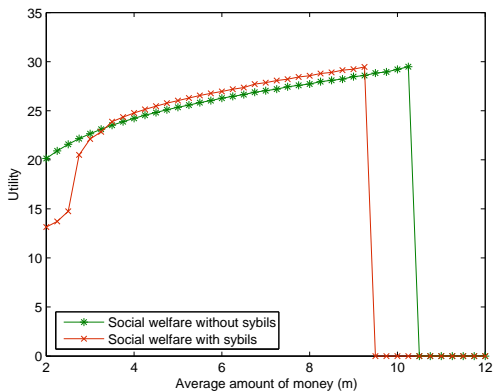}
\caption{Sybils can cause a crash}
\label{fig:sybilcrash}
\end{figure}

We have seen that when a large fraction of agents have sybils, those
agents without sybils tend to be starved of opportunities to work
(i.e. they have a low value of $p_u$).
However, as Figure~\ref{fig:sybilcost} shows, when a small
fraction of agents have sybils this effect (and its corresponding
cost) is small.  Surprisingly, if there are few agents with sybils,
an increase in the number of sybils these agents have can actually
result in a \emph{decrease} of their effect on the other agents.
Because agents with sybils are more likely to be chosen to satisfy any
particular request, they are able to use lower thresholds and reach
those thresholds faster than they would without sybils,
so fewer are competing to satisfy any given
request.  Furthermore, since agents with sybils can almost always pay
to make a request, they can provide more opportunities for other agents
to satisfy requests
and earn money.  Social welfare is essentially proportional to
the number of satisfied requests (and is exactly proportional to it if
everyone shares the
same values of $\alpha$ and $\gamma$), so a small number of agents
with a large number of sybils can improve social welfare, as
Figure~\ref{fig:sybilsmall} shows.
Note that, although social welfare increases, some agents may be worse
off.  For example,
for the choice of parameters
in this example, social welfare increases when
twenty percent of agents create at least two sybils, but agents
without sybils are
worse off unless the twenty percent of agents with sybils create at
least eight sybils.
As the number of agents with sybils increases, they
start competing with each other for opportunities to earn money and so
adopt higher thresholds, and this benefit disappears.
This is what causes the discontinuity in Figure~\ref{fig:sybilcost} when
approximately a third of the agents have sybils.

This observation about the discontinuity also suggests another way to
mitigate the negative effects of sybils:
increase the amount of money in the system.
This effect can be seen in Figure~\ref{fig:sybilcrash}, where for $m =
2$ social welfare is very low with sybils but by $m = 4$ it is higher
than it would be without sybils.

Unfortunately, increasing the average amount of money has its own
problems.
Recall from Section~\ref{sec:optimize} that, if the average amount of
money per agent is too high, the system will crash.
It turns out than just a small number of agents creating
sybils can have the same effect, as
Figure~\ref{fig:sybilcrash} shows.
With no sybils, the point at which social welfare stops increasing and
the system crashes is between $m = 10.25$ and $m = 10.5$
(we only calculated social welfare for values of $m$ that are
multiples of 0.25, so we do not know the exact point of the crash).
If one-fifth of the agents each create a single sybil, the
system crashes if $m=9.5$,
a point where, without sybils, the social welfare was near optimal.
Thus, if the system designer tries to induce optimal behavior without
taking sybils into account, the system will crash.
Moreover, because of the possibility of a crash,
raising $m$ to tolerate more sybils is effective only if $m$
was already set conservatively.

This discussions shows that the presence of sybils can have a
significant impact on the tradeoff between efficiency and stability.
Setting the money supply high can increase social welfare, but at the
price of making the system less stable. Moreover, as the following
theorem shows, whatever
efficiencies can be achieved with sybils can be achieved without
them, at least if there is only one type of agent.
In the theorem, we
consider a system where all agents have the same type $t$.
Suppose that some subset of the agents have created sybils, and all the
agents in the subset have created the same number of sybils.  We can
model this by simply taking the agents in the subsets to have a new type
$s$, which is identical to $t$ except that the value of $\chi$ increases.
Thus, we state our results in terms of systems with two types of agents,
$t$ and $s$.

\thm \label{thm:equivalent}
Suppose that $t$ and $s$ are two types that agree except for the value
of $\chi$, and that $\chi_t < \chi_s$.
If $\vec{k} = (k_t,k_s)$ is an $\varepsilon$-Nash equilibrium for
$G = (\{t,s\},\vec{f},h,m,n)$ with social welfare $w$, then there exist
$h'$, $m'$, and $n'$ such that $\vec{k}' = (k_s)$ is an $\varepsilon$-Nash
equilibrium for $G'_{h',m',n'} = (\{ t \}, \{ 1 \},h',m',n')$ with
social welfare
greater than $w$.
\ethm

We defer proof of Theorem~\ref{thm:equivalent} to
Appendix~\ref{sec:nonstandard_proofs}.

The analogous result for systems with more than one type of agent is
not true.  Figure~\ref{fig:sybilcost} shows a game with a single type
of agent, some of whom have created two sybils.  However, we can
reinterpret it as a game with two types of agents, one of whom
has a larger value of $\chi$.  With this reinterpretation,
Figure~\ref{fig:sybilcost} shows that social welfare is higher when
all the agents are of the type $t_h$ with the higher value of $\chi$
than when only 40\% are.  Moreover, if only 40\% of the agents have type
$t_h$, social welfare would increase if the
remaining agents created two sybils each (resulting in all
agents having the higher value of $\chi$).
Note that this situation, where there are two types of agents, of which
one has a higher value of $\chi$, is exactly the situation considered by
Theorem~\ref{thm:equivalent}.  Thus, the theorem shows that for any
equilibrium with two such types of agents, there is a better equilibrium
where one of those types creates sybils so as to effectively create only
one type
of agent.

While situations like this
show that it is theoretically possible for sybils to increase social
welfare beyond what is possible to achieve by simply adjusting the
average amount of
money, this outcome seems unlikely in practice. It relies on
agents creating just the right number of sybils. For situations
where such a precise use of sybils would lead to a significant
increase in social welfare, a designer could instead improve social
welfare by biasing the algorithm agents use for selecting which
volunteer will satisfy the request.

Thus far, we have assumed that when agents create sybils the amount of
money in the system does not change.  However, the presence of sybils
increases the number of apparent agents in the system.  Since social
welfare depends on the average amount of money
per agent,
if the system designer
mistakes these sybils for an influx of new users and increases the
money supply accordingly, she will actually end up increasing the
average amount of money in the system,
and may cause a crash.
This emphasizes the need for
continual monitoring of the system rather that just using simple
heuristics to set the average amount of money, an issue we discuss
more in Section~\ref{sec:identify}.

\subsection{Collusion}

Agents that collude gain two major benefits.  The primary benefit is
that they can share money,%
\footnote{We assume that colluding agents act to maximize the sum of
their
  utilities.  Of course, this may not be optimal for any particular
  agent, so sustaining collusion is a problem for would-be colluders.}
which makes them less likely
to run out of money (and hence unable to make a request), and allows them to
pursue a joint strategy for determining when to work.  A secondary
benefit, but important in particular for larger collusive groups, is
that they can satisfy each other's requests.  The effects of collusion
on the rest of the system depend crucially on whether agents are
able to volunteer to satisfy requests when they personally cannot
satisfy the request but one of their colluding partners can.  In a
system where a request is for computation, it seems relatively
straightforward for an agent to pass the computation to a partner to
perform and then pass the answer back to the requester.  On the other
hand, if a request is a piece of a file it seems less plausible that
an agent would accept a download from an unexpected source,
and it seems wasteful to have the chosen volunteer download it
for the sole purpose of immediately uploading it.  If it is possible
for colluders to pass off requests in this fashion, they are able to
effectively act as sybils for each other, with all the consequences
discussed in Section~\ref{sec:sybil}.  However, if agents can
volunteer only for requests they can personally satisfy, the effects of
collusion are almost entirely positive.

Since we have already discussed the consequences of sybils, we will
assume that agents are able to volunteer only to satisfy requests that
they personally can satisfy.
Furthermore, we make the simplifying assumption that agents that
collude are of the same type, because if agents of different types
collude their strategic decisions become more complicated.  For
example, once the colluding group has accumulated a certain amount of
money, it may wish to have only members with small values of
$\alpha$ volunteer to satisfy requests; or when it is low on money, it
may wish to deny use of money to members with low values of $\gamma$.
This results in strategies that involve sets of thresholds rather than
a single threshold. While there seems to be nothing fundamentally different
about the situation, it makes calculations significantly more difficult.

With these assumptions, we now examine how colluding agents will
behave.
Because colluding agents share money and types, it is irrelevant which
members actually perform work and have money.  All that matters is the
total amount of money the group has.  This means that when the group
needs money, everyone in the group volunteers for a job; otherwise, no
one does.  Thus, the group essentially acts like a single
agent, using a threshold
that is
somewhat less than the sum of the thresholds
that the individual agents
would have used,
because it is less likely that $c$ agents will make $ck$ requests in
rapid succession than a single agent making $k$.  Furthermore, some
requests will not require scrip at all because they can potentially be
satisfied by other members of the colluding group.
When deciding whether the
group should satisfy a member's request or ask for an outside
volunteer to fulfill it, the group must decide whether it should pay a
cost of $\alpha$ to avoid spending a dollar.  Since not spending a
dollar is effectively the same as earning a dollar, the decision is
already optimized by the threshold strategy; the group should always
attempt to satisfy a request internally unless it is in a temporary
situation where the group is above its threshold.

\begin{figure}[htb]
\centering \epsfig{file=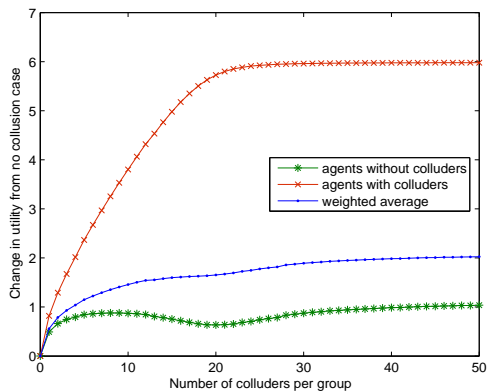, height=2.2in}
\caption{The effect of collusion on utility}
\label{fig:numcollusion}
\end{figure}

Figure~\ref{fig:numcollusion} shows an example of the effects of
collusion on agents' utilities as the size of collusive groups
increases.  As this figure suggests, the effects typically go through
three phases.  Initially, the fraction of requests colluders satisfy
for each other is small.  This means that each collusive group must
work for others to pay for almost every request its members make.
However, since they share money, the colluders do not have to work as
often as individuals would.  Thus, other agents have more
opportunity to work, and every agent's $p_u$ increases,
making all agents better off.

As the number of colluders increases,
the fraction of requests they satisfy internally grows significant.
We can think of $p_d$ as decreasing in this case, and view these
requests as being satisfied ``outside'' the scrip system because no
scrip changes hands.
This is good for colluders,
but is bad for other agents whose $p_u$ is lower,
since fewer requests are being made.  Even in this range,
non-colluding agents still tend to be better off than if
there were no colluders, because the overall competition for
opportunities to work is still lower.  Finally, once the collusive
group is large enough, it will have a low $p_d$ relative to $p_u$.
This means the collusive group can use a very low threshold, which
again begins improving utility for all agents.
The analogous
situation with sybils is transitory, and disappears when more agents
create sybils.  However, with collusion,
this low threshold is an inherent
consequence of colluders satisfying each other's requests, and so
persists and even increases as the amount of collusion in the system
increases.
Since collusion is difficult to maintain (the problem of incentivizing
agents to contribute is the whole point of using scrip), we would
expect the size of collusive groups seen in practice to be relatively
small.  Therefore, we expect that for most systems collusion
will make no agent worse off, and some better off.  Note that, as with
sybils, the decreased in
competition that results from collusion can also lead to a crash.
However, if the system designer
is monitoring the system, and encouraging and expecting collusion, she
can reduce $m$ appropriately and prevent a crash.

These results also suggest that creating the
ability to take out loans (with an appropriate interest rate) is
likely to be beneficial.
Loans gain the benefits of reduced competition without the
accompanying cost of fewer requests being made in the system.
However, implementing a loan mechanism requires addressing a number of
other incentive problems.  For example, {\em whitewashing}, where
agents take on a new identity (in this case to escape debts) needs to
be prevented~\cite{FrR01}.

\section{Identifying User Strategies} \label{sec:identify}

Lemma~\ref{lem:minrelent} used
relative entropy to derive
an explicit formula for the distribution of money $d^*$ given
a game $(T,\vec{f},h,n,m)$ and vector of strategies $\vec{k}$.
In this section, we want to go in the opposite direction:
given the distribution of money, we want to infer the strategies
$\vec{k}$, the set of types present $T$, and the fraction of each type
$\vec{f}$.
For those interested in understanding the agents using a scrip system,
knowing the fraction of agents using each strategy can provide a window
into the preferences of those agents.
For system designers, this knowledge is useful because, as we
show in Section~\ref{sec:optimize}, how much money the system can
handle without crashing depends on the fraction of agents of each
type.

In equilibrium, the distribution of money has the form
described in Lemma~\ref{lem:minrelent}.
Note that, in general, we do not expect to see exactly this
distribution at any given time,
but it follows from Theorem~\ref{thm:distribution}
that, after sufficient time,  the distribution is unlikely
to be very far from it.
Does this distribution help us identify the strategies and types of
agents?

As a first step to answering this question, given a distribution of
money $d$ (where $d(i)$ is the fraction of agents with $i$ dollars) such
that $d(i)$ is
a rational number
for all $i$ (this constraint is necessary if $d(i)$
is to represent the fraction of agents with $i$ dollars in a real system),
suppose that the maximum
amount of money to which $d$ gives positive probability is $K$.
A vector $\vec{f}$ of length $K+1$ whose components are all rational
numbers, where $f_i$ is intuitively the
fraction of agents playing the threshold strategy $s_i$, is an
{\em explanation} of $d$ if there exists a $\lambda$ such that
$$d(j) = \sum_i d_{\lambda}(i,j),$$ where
\begin{equation}
\label{eqn:explanation}
d_{\lambda}(i,j) = f_i \lambda^j / (\sum_{l = 0}^i \lambda^l)
\end{equation}
if $j \leq i$ and 0 otherwise.
Note that Equation~(\ref{eqn:explanation}) is very similar to
Equation~(\ref{eqn:d}) from Lemma~\ref{lem:minrelent}.  In
the following lemma, we show why we call $\vec{f}$ an explanation:
given a distribution $d$ and an explanation $\vec{f}$ we can find a
game $G$ where $\vec{f}$ is the fraction of agents of each type and
$d$ is the equilibrium distribution of money (by which
we mean that the
value of $d^*$ in Lemma~\ref{lem:minrelent} is such that
$d(i) = \sum_{t} d^*(t,i)$).
Note that this definition implicitly assumes that $w_t = 1$ for all
$t$, a point to which we return later.

\lem \label{lem:strategies}
If $\vec{f}$ is an explanation for $d$,
then there exists a game
$G = (T,\vec{f},h,m,n)$ and vector $\vec{k}$ of thresholds such that
$\vec{k}$ is an $\varepsilon$-Nash
equilibrium for $G$ and the equilibrium distribution of
money is $d$.
\elem

\prf
Let $T = \{ 0, \ldots, K \}$,
$h$ be the minimum integer such that $hd(i)$ is an integer for all
$i$,
$m = \sum_i i d(i)$,
and
$\vec{k}$ be such that $k_i = i$.
For each type $i$, choose $\beta_i$, $\chi_i$, and $\rho_i$ arbitrarily,
subject to the constraint that  $\beta \chi / \rho = 1$
(so that, by definition, $\omega_i = 1$ for all types $i$).
Finally, choose an arbitrary $n$.

By Lemma~\ref{lem:br}, for any $n$, an optimal threshold policy in the MDP
$\mathcal{P}_{G,\vec{S}(\vec{k}),i}$ for an agent of type $i$ is
$s_\kappa$, where $\kappa$ is the maximum value such that
\begin{equation}
\tag{\ref{eqn:policy}}
\alpha_t \leq E[(1 - (1 - \delta_t)/n)^{J(\kappa,p_u,p_d)}] \gamma_t.
\end{equation}
Fix $\delta_i$ and $\gamma_i$, and
let $g(\kappa)$ be the sequence of values of the right hand side of
Equation~(\ref{eqn:policy}) for natural numbers $\kappa$.
Recall that the random variable $J(\kappa,p_u,p_d)$ represents the round at
which an agent
starting with $\kappa$ dollars runs out of money.
Since $J(0,p_u,p_d) = 0$ for all histories, $g(0) = \gamma_t$.
The time at which an agent runs out of money is increasing in his
initial amount of money.  Thus,
$J(\kappa,p_u,p_d)$ is a strictly
increasing
function of $\kappa$,
so $g(\kappa)$
is strictly decreasing.
Choose $\alpha_i$ such that $g(i+1) < \alpha_i < g(i)$.

Thus, we have established parameters
$(\alpha_i,\beta_i,\gamma_i,\delta_i,\chi_i,\rho_i)$ for each type $i$
so that $s_i$ an optimal policy for agents of type $i$ in the MDP
$\mathcal{P}_{G,\vec{S}(\vec{k}),i}$.
By Theorem~\ref{thm:threshold}, taking $n$ and the $\delta_i$
sufficiently large makes $\vec{k}$ a $\varepsilon$-Nash equilibrium
for $G$.  By Lemma~\ref{lem:minrelent}, the equilibrium distribution
of money is $d$.
\eprf

In general, there is not a unique explanation of a distribution $d$.
Say that a distribution of money $d$
is \emph{fully-supported} if there do not exist
$i$ and $j$ such that $i < j$, $d(j) > 0$, and $d(i) = 0$.
For any game $G$, if all agents play threshold strategies then the
resulting distribution will be fully-supported because it has the form
given in Lemma~\ref{lem:minrelent}.
As the following lemma shows,
a fully-supported distribution can be explained in an infinite
number of different ways.

\lem \label{lem:infinite}
If $d$ is a fully-supported distribution of money with finite support,
then there exist an infinite number of explanations of $d$.
\elem

We defer the proof of Lemma~\ref{lem:infinite} to
Appendix~\ref{sec:identify_proof}.

Lemma~\ref{lem:infinite} shows that $d$ has an infinite number of
explanations.  Lemma~\ref{lem:strategies} shows that we can find an
(approximate)
equilibrium corresponding to each of them.
The explanations $\vec{f}$ we construct in the proof of
Lemma~\ref{lem:infinite} seem unnatural;
typically $f_i > 0$ for all $i$.
We are interested in a more parsimonious explanation, one that has a
small support (i.e., the number of thresholds $i$ for which $f_i > 0$ is
small),
for reasons the following lemma makes clear.

\lem \label{lem:uniquemoney}
Let $\vec{f}$ be an explanation for $d$.  If
$s$ is the size of the support of
$\vec{f}$, then any other explanation will have a support of size
at least $K - s$.
\elem

\prf
Suppose that $\vec{f}$ is an explanation for $d$.
By Lemma~\ref{lem:strategies}, there is a game $G =
(T,\vec{f},h,m,n)$ and vector $\vec{k}$ of thresholds such that
$\vec{k}$ is an $\epsilon$-Nash equilibrium for $G$ and the equilibrium
distribution of money is $d$.  Moreover, the proof of
Lemma~\ref{lem:strategies} shows that we can take $T = \{0,\ldots,K\}$,
$k_i = i$, and $\omega_i = 1$ for each type $i \in T$.
By Equation~(\ref{eqn:d}) in Lemma \ref{lem:minrelent},
$d^*(t,i) = f_t \lambda^i q(t,i) /
\sum_{j = 0}^{k_t} \lambda^j q(t,j)$,
where $\lambda$ is the (unique) value that satisfies
Equation~(\ref{eqn:m}).
We first show that if $f_{i-1} = 0$, then $d(i) / d(i-1) = \lambda$.
Since, for all $i$, $\omega_i = 1$,
it is immediate from
the definition of $q$ for Lemma~\ref{lem:minrelent} that
$q(i,j) = q(i,j')$ for all $j$ and $j'$.
Thus, the $q$ terms cancel, so
$d^*(i,j) = f_i \lambda^j / \sum_{l = 0}^{k_i} \lambda^l$.
Let $b_i = f_i / \sum_{l = 0}^{k_i} \lambda^l$;
then $d^*(i,j) = \lambda^j b_i$.
Only agents with a threshold of at least $j$ can have $j$ dollars, so
$$d(j) = \sum_j d^*(i,j) = \sum_{\{t_l: l \ge j\}} d^*(l,j)
= \sum_{\{t_l: l \ge j\}} b_{l} \lambda^j = B_j \lambda^j,$$
where $B_j = \sum_{\{t_l: l \ge i\}} b_{l}$.
If $f_{{i-1}} = 0$,  then $B_i =
B_{i-1}$, so $d(i) / d(i-1) = \lambda$.

Since $s$ strategies get positive probability according to $\vec{f}$,
at least $k - s$ of the ratios $d(i)/d(i-1)$ with $1 \le i \le K$
must have value $\lambda$.
Any other explanation $\vec{f}'$ will have
different coefficients $f_i$ in Equation~(\ref{eqn:explanation}), so
the value
$\lambda'$ satisfying it will also differ
(since the requirement that $d(K) = d_{\lambda}(K,K)$ uniquely defines a
value of $\lambda$).
This means that the $K - s$ ratios with value $\lambda$ must
correspond to strategies $i$ such that $f_i > 0$.
Thus, the support of any other
explanation must be at least $K - s$.
\eprf

If $s \ll K$, Lemma \ref{lem:uniquemoney} gives us a strong reason
for preferring the minimal explanation (i.e., the one with the smallest
support); any other explanation will
involve significantly more types of agents being present.
For $s = 3$ and $K = 50$,
the smallest explanation has a support of three thresholds, and
thus requires three types;
the next smallest explanation requires
at least
47 types.  Thus, if the number of types
of agents is relatively small, the minimal explanation will be the
correct one.

\begin{figure}[htb]
\centering \epsfig{file=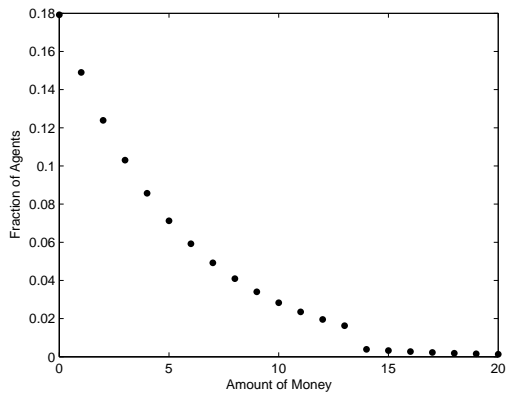, height=2.2in}
\caption{Distribution of money with two types of agents.}
\label{fig:distribution}
\end{figure}

\begin{figure}[htb]
\centering \epsfig{file=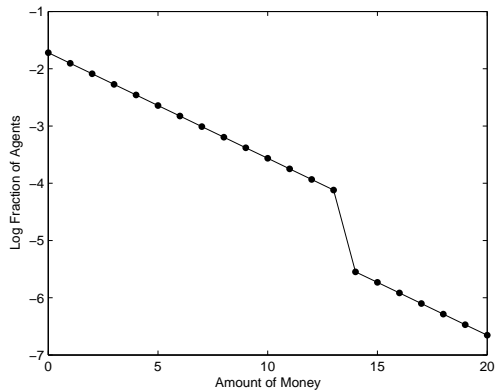, height=2.2in}
\caption{Log of the distribution of money with two types of agents.}
\label{fig:log}
\end{figure}

The proof of Lemma~\ref{lem:uniquemoney} also gives us an algorithm
for finding this minimal explanation. Since $d(i) = B_i
\lambda^i$, taking logs of both sides, $\log d(i) = \log B_i + i \log
\lambda$.
Because $B_i$ is constant in ranges of $i$
where $f_{t_i} = 0$,
a plot of $\log d(i)$ will be a line with slope $\lambda$ in these
ranges.
Thus, the minimal
explanation can be found by finding the minimum number of lines of
constant slope that fit the data.  For a simple example of how such a
distribution might look, Figure~\ref{fig:distribution} shows an
equilibrium distribution of money for the game
$$(\{(.05,1,1,.95,1),(.15,1,1,.95,1) \},(.3,.7),10,4,100)$$
so the only
difference between the types is that it costs the second type three times
as much to satisfy a request) and the equilibrium
strategy profile $(20,13)$.
Figure~\ref{fig:log} has the same distribution plotted on a log
scale.  Note the two lines with the same slope ($\lambda$) and the
break at 13.

Our notion of an explanation requires that $\vec{f}$ satisfy
Equation~(\ref{eqn:explanation}), which, unlike
Equation~(\ref{eqn:d}), does not contain a $q(t,i)$ term.  Thus, it
implicitly assumes that for all types $t$, $\omega_t = 1$.
Note that the game $G$ in the proof of
Lemma~\ref{lem:uniquemoney} was constructed so at to ensure that
$\omega_t = 1$ for all types $t$.
When $\omega_t$ is
allowed to differ, we no longer have the simple form for $B_i$
used in Lemma~\ref{lem:uniquemoney}.  This is because the types do not
share the same value of $\omega_t$.
However,
for a single type $t$ it is the case that
$d^*(t,i) / d^*(t,i-1) = \lambda \omega_t$.
$\omega_{\tau(j)}$ can be estimated by observing the results of
requests, so by
observing a sufficient number of agents the system designer should be
able to estimate the values $d^*(t,i)$ and $\omega_t$ for some type
$t$ and thus learn $\lambda$.
If several, but not all, types $t$
have a common value of $\omega_t$, the procedure above can be used to
determine $f_t$ and $k_t$ for each type and the resulting
value of $\lambda$.

This procedure allows us to use a distribution of money to infer the
minimal explanation of the
number of types of agents: the fraction of
the population composed of each type, and the strategy each type is
playing. (Note that we cannot distinguish multiple types with a shared
$\omega_t$ playing the
same strategy.)  We would like to use this information to learn about
the preferences of agents: their values of $\alpha_t$, $\gamma_t$, and
$\delta_t$.  Lemma~\ref{lem:br} shows how we can do this.  Once we find
an explanation, the value of $\lambda$ determines $p_u^t$ and $p_d^t$ for
each type $t$.  Then Equation~\ref{eqn:policy} puts constraints on the
values of $\alpha_t$, $\beta_t$, and $\gamma_t$.
Over time, if $T$, the set of types,
remains constant, but $\vec{f}$, $n$, and $m$ all vary as
agents join and leave the system,  a later observation with a slightly
would give another equilibrium
with new constraints on the types of the agents.  A number of
such
observations potentially reveal enough information to
allow strong inferences about agent types.

Thus far we have implicitly assumed that there are only a small number
of types of agents in a system.
Given that a type is defined by six real numbers, it is perhaps more
reasonable to assume that each agent has a different type, but there is
a small number of ``clusters'' of agents with
similar types.
For example, we might believe that generally agents either place a
high value or a low value on receiving service.
While the exact value may vary, the types of two low-value agents
or two high-value agents will be quite similar.
We have also assumed in our analysis that all agents
play their optimal threshold strategy.  However, computing this
optimum may be too difficult for many agents.
Even ignoring computational issues, agents may have
insufficient information about their exact type
or the exact types of other agents to compute the optimal threshold strategy.
Both the assumption that there are a few clusters of agents with
similar, but not
identical, types and the assumption that agents do not necessarily play
their optimal threshold strategy, but do play a strategy close to
optimal, lead to a similar picture of a system, which is one that we
expect to see in practice:
we will get clusters of agents playing similar strategies (that is,
strategies with thresholds clustered around one value), rather than all
agents in a cluster playing exactly the same strategy.
This change has relatively little impact on our results.
Rather than seeing straight lines representing populations with a
sharp gap between them, as in Figure~\ref{fig:log}, we expect slightly
curved lines representing a cluster of similar populations,
with somewhat smoother transitions.

\section{Discussion} \label{sec:conclusion}

In this paper, we have examined some
of the practical implications of the the theoretical results about
scrip systems from our companion paper~\cite{scripjournal1}.  For
those interested in studying the agents of scrip systems, our
characterization of equilibrium distribution of money forms the basis
for techniques relevant to inferring characteristics of the agents of a
scrip system from the distribution of money.   For a system designer,
our results on optimizing the money supply provide a simple maxim:
keep adding money until the system is about to experience a monetary
crash.

We have also seen that our model can be used to understand the effects
of nonstandard agent behavior on a scrip system.
It provides insight into the effects of altruists and hoarders on a scrip
system and guidance to system designers for dealing
with them (less and more money respectively).
Sybils are generally bad,
but can typically be discouraged by imposing a moderate cost and
possibly biasing the process for selecting a volunteer.
On the other hand, collusion tends to be a net benefit and
should be encouraged.  Indeed, the entire purpose of the system is to
allow users to collude and provide each other with service despite
incentives to free ride.

We remark that we are not the first to study the
effects of altruists, sybils, and collusion on system
behavior (although we believe we are the first to study it in the
context of scrip systems).
Work on the evolution of cooperation stresses the importance of
altruists willing to undertake costly punishment~\cite{Nowak07}.
Yokoo et al.~\citeyear{yokoo04} studied the effects of sybils in
auctions.
Solution concepts such as \emph{strong Nash equilibrium}~\cite{strong}
and \emph{$k$-$t$ robust equilibrium}~\cite{ADGH06} explicitly address
collusion in games; Hayrapetyan et al.~\citeyear{hayrapetyan06} study
collusion in
congestion games and find cases where, as with scrip systems,
collusion is actually beneficial.

Although we believe that our analysis should already provide a great
deal of insight to a system designer hoping to use a scrip system,
many interesting open questions remain for future work.  To
name a few:
\begin{itemize}

\item
Our model makes a number of strong predictions about the agent
strategies, distribution of money, and effects of variations in the
money supply.  It also provides techniques to help analyze
characteristics of agents of a scrip system.  It would be interesting
to test these predictions on a real functioning scrip system to either
validate the model or gain insight from where its predictions are
incorrect.

\item In many systems
there are overlapping communities of various sizes that are
significantly more likely to be able to satisfy each other's
requests.
For example, in a P2P filesharing system, people are more likely to be
able to satisfy the requests of others who share the same interests.
It would be interesting to investigate the effect of such communities on
the equilibrium of our system.

\item
It seems unlikely that altruism and hoarding are the only two
types of ``irrational'' behavior we will find in real systems.  Are there
other major types that our model can provide insight into?
Furthermore,
it seems natural that the behavior of a very small group of agents
should not be able to change the overall behavior of the system.  Can
we prove results about equilibria and utility when a small group
follows an arbitrary strategy?  This is particularly relevant when
modeling attackers.  See \cite{ADGH06} for general results in this
setting.

\end{itemize}

\section{Acknowledgments}

We would like to thank
Randy Farmer,
Peter Harremoes,
Shane Henderson,
Jon Kleinberg,
David Parkes,
Dan Reeves,
Emin G\"{u}n Sirer,
Michael Wellman,
and anonymous referees
for helpful suggestions, discussions, and criticisms.
Eric
Friedman, Ian Kash, and Joseph Halpern are supported in part by NSF
under grant
ITR-0325453.
Joseph Halpern is
also supported in part by
NSF under grants CTC-0208535, IIS-0812045, and IIS-0534064;
by ONR under grant N00014-01-10-511;
by the DoD Multidisciplinary University Research Initiative (MURI)
program administered by the ONR under grants N00014-01-1-0795 and
N00014-04-1-0725;
and by AFOSR under grants F49620-02-1-0101, FA9550-08-1-0438,
FA9550-05-1-0055,
and FA9550-09-1-0266.
Eric Friedman is also supported in part by NSF under grant CDI-0835706.

\appendix

\section{Definition of the MDP}
\label{sec:MDP}

\sloppypar{
In this appendix, we repeat the formal definition of the MDP from our
companion paper~\cite{scripjournal1}.
Taking notation from Puterman~\citeyear{puterman},
we formally define the MDP $\mathcal{P}_{G,\vec{S}(\vec{k}),t}
= \mbox{$(S,A,p(\cdot \mid s,a),r(s,a))$}$ that describes the game where
all the
agents other than $i$ are playing $\vec{S}(\vec{k})_{-i}$ and $i$ has
type $t$.
}
\begin{itemize}
\item $S =
\{0, \ldots , mhn \}$ is the set of possible states for the MDP (i.e., the
possible amounts of money compatible with the distribution $d^*$).
\item $A = \{0 , 1 \}$ is the set of possible actions
for the agent,
where 0 denotes not volunteering and 1 denotes volunteering
iff another agent who has at least one dollar makes a request.
\item $p_u$ is the probability of earning a dollar, assuming the agent
volunteers
(given that all other agents have fixed their thresholds according to
$\vec{k}$ and the distribution of money is exactly
$d^*$.  Each agent of type $t'$ who wishes to volunteer can do so with
probability $\beta_{t'}$.  Assuming exactly the expected number of
agents are able to volunteer,
$\upsilon_{t'} = \beta_{t'}(f_{t'} - d^*(t',k_{t'}))n$
agents of type $t'$ volunteer.  Note that we are disregarding the
effect of $i$ in computing the $\upsilon_{t'}$, since this will have a
negligible effect for large $n$.
Using the $\upsilon_t$s, we can express $p_u$ as the product of two
probabilities: that some agent other than $i$ who has a dollar is
chosen to make a request and that $i$ is the agent chosen to satisfy
it. Thus,
\begin{equation}
\label{eqn:pu}
p_u = \left(\sum_{t'} \rho_{t'}( f_{t'} - d^*(t',0))\right)
\left(\frac{\chi_t\beta_t}{\sum_{t'}\chi_{t'}\upsilon_{t'}}\right).
\end{equation}
\item $p_d$ is the probability of agent $i$ having a request satisfied, given
that agent $i$ has a dollar.
Given that all agents are playing a threshold strategy, if the total
number $n$ of agents is sufficiently large, then it is almost certainly
the case that some agent will always be willing and able
to volunteer.  Thus, we can take $p_d$ to be the probability that agent
$i$ will be chosen to make a request; that is,
\begin{equation}
\label{eqn:pd}
p_d = \frac{\rho_t}{hn}
\end{equation}
\item $r(s,a)$ is the (immediate) expected reward for performing
action $a$ in state $s$.  Thus, $r(s,0) = \gamma_t p_d $ if $s > 0$;
$r(0,0) = 0$; $r(s,1) = \gamma_t p_d - \alpha_t p_u$ if $s > 0$; and
$r(0,1) = - \alpha_t p_u$.
\item $p(s' \mid s,a)$ is the probability of being in state $s'$ after
performing action $a$ in state $s$;
$p(s' \mid s,a)$ is determined by $p_u$ and $p_d$; specifically,
$p(s+1 \mid s,1) = p_u$, $p(s-1 \mid s,a) = p_d$ if $s > 0$, and
the remainder of the probability is on $p(s \mid s,a)$
(i.e., $p(s \mid s,a) = 1 - (p(s+1 \mid s,1) + p(s-1 \mid s,a)$).
\item $u^*(s)$ is the expected utility of being in state $s$
if agent $i$ uses the optimal policy for the MDP
$\mathcal{P}_{G,\vec{S}(\vec{k}),t}$
\item $u(s,a)$ is the expected utility for performing action $a$ in
state $s$, given that the optimal strategy is followed after this
action;
$$u(s,a) = r(s,a) + \delta \sum_{s'=0}^{mhn} p(s' \mid s,a) u^*(s').$$
\end{itemize}

\section{Proofs from Section~\ref{sec:nonstandard}}
\label{sec:nonstandard_proofs}

\rethm{thm:peps}
Fix a game $G$ and vector of thresholds $\vec{k}$.
Let $R_{\vec{k},t} = p_u^{\vec{k},t} / p_d^t$.
In the limit as the number of rounds goes to
infinity, the fraction of the agent's requests that have an agent
willing and able to satisfy them that get satisfied is
$(R_{\vec{k},t} - R_{\vec{k},t}^{k_t+1}) / (1 - R_{\vec{k},t}^{k_t+1})$
if $R_{\vec{k},t} \neq 1$ and
$k_t / (k_t + 1)$ if $R_{\vec{k},t} = 1$.
\erethm

\prf
Consider the Markov chain $\mathcal{M}$
that results from fixing the agent's policy to $s_{k_t}$ in
$\mathcal{P}_{G,\vec{S}(\vec{k}),t}$.
$\mathcal{M}$ satisfies the requirements
to have a limit distribution (see Theorem A.1 of~\cite{scripjournal1}).
It can be easily verified that the distribution
gives the agent  probability $R^i (1 - R) / (1 - R^{k+1})$ of having $i$
dollars if $R \neq 1$ and probability $1/(k+1$) if
$R = 1$ satisfies the detailed balance condition and thus is the limit
distribution.  This gives the probabilities given in the theorem.
\eprf

\rethm{thm:equivalent}
Suppose that $t$ and $s$ are two types that agree except for the value
of $\chi$, and that $\chi_t < \chi_s$.
If $\vec{k} = (k_t,k_s)$ is an $\varepsilon$-Nash equilibrium for
$G = (\{t,s\},\vec{f},h,m,n)$ with social welfare $w$, then there exist
$m'$, and $n'$
such that $\vec{k}' = (k_s)$ is an $\varepsilon$-Nash
equilibrium for $G'_{m',n'} = (\{ t \}, \{ 1 \},h,m',n')$
with social welfare greater than $w$.
\erethm

\prf
We prove the theorem by finding $m'$, and $n'$ such that agents
in $G'_{m',n'}$
that play some strategy $k$ get essentially the same utility that an
agent with sybils would by playing that strategy in $G$.
Since $k_s$ was the optimal strategy for agents with sybils in $G$,
it must be optimal in $G_{m',n'}$ as well.
Since agents with sybils have utility at least as great as those
without, social welfare will be
at least as large in $G'_{m',n'}$ as in $G$.
To do so, we find a value of $m'$ so that, from his perspective, being
in $G$ with sybils or $G'$ without results in exactly the same MDP. The
natural way to do so is to treat $m'$ as a continuous value, which
might result in a value such that $hm'n$ (the total amount of money)
is not an integer. To complete the proof, we show that an $n'$ can be
found that allows us to avoid this problem.

Since an agent can earn a dollar only if he is able to satisfy the
current request, $0 < p_u^{m,\vec{k},s} < \beta_s$.
The constraint that $hm'n'$, the total amount
of money, is a natural number means that
$m'$ must be a rational number.  For the moment, we
ignore that constraint and allow $m'$ to take on any value in
$[0,k_t']$.
From Equation~(\ref{eqn:pu}), $p_u^{m',\vec{k}',t}$ is continuous in
$d^*_{q_{\vec{k}'}}$, which, by Lemma~\ref{lem:minrelent},
is continuous in $\lambda_{m',\vec{k}'}$ and thus $m'$.
We use this continuity to show that we can find a value of $m'$ such
that $p_u^{m,\vec{k},s} = p_u^{m',\vec{k}',t}$.
By Equation~(\ref{eqn:m}), if $m' = 0$ then
$d^*_{q_{\vec{k}'},m}(t,0) = 1$, and if $m' = k_t'$ then
$d^*_{q_{\vec{k}'},m}(t,k_t') = 1$.
Combining these with Equation~(\ref{eqn:pu}) gives
$p_u^{0,\vec{k}',t} = 0$ and
$p_u^{m',\vec{k}',t} = \beta_t.$
Thus, by the Intermediate Value Theorem, there exists an $m'$ such that
$p_u^{m,\vec{k},s} = p_u^{m',\vec{k}',t}$.
%
For this choice of $m'$, observe that by Lemma~\ref{lem:br},
$\mathcal{P}_{G,\vec{S}(\vec{k}),s}$ and
$\mathcal{P}_{G_{m'},\vec{S}(\vec{\kappa}'),t}$ have the same optimal
threshold policy.

If $m'$ is rational, say $m' = a/b$, take $n' = bn$; then $hm'n'$ is
an integer and, by the argument above,
$\vec{k}'$ is an equilibrium for $G_{m',n'}$.
Since $p_u^{m',\vec{k}'} = p_u^{\vec{k},s}
> p_u^{\vec{k},t}$, we must have $\zeta_{G_{m',n'}} < \zeta_G$.   Thus,
social welfare has increased.
If $m'$ is not rational, we instead use a rational value $m''$
sufficiently close to $m'$ that $\vec{k}'$ is still an equilibrium for
$G_{m',n'}$ and $\zeta_{G_{m'',n''}} < \zeta_G$.
\eprf

\section{Proof of Lemma~\ref{lem:infinite}} \label{sec:identify_proof}

\relem{lem:infinite}
If $d$ is a fully-supported distribution of money with finite support,
there exist an infinite number of explanations of $d$.
\erelem

\prf
Fix $\lambda$.  The distribution $d$ and $\lambda$ determine an
explanation $\vec{f}$ as follows.
By Equation~(\ref{eqn:explanation}), we need $\vec{f}$ to satisfy
$d(j) = \sum_{i = 0}^K d_{\lambda}(i,j)$.

Recall that $K$ is the maximum value for which $d(K) > 0$.
Start by considering $f_K$.
By the definition of $d_{\lambda}$, $d_{\lambda}(i,j) = 0$
if $j > i$.  Thus, the constraint becomes
$$d(K) = d_{\lambda}(K,K) = f_{K} \lambda^K
/ (\sum_{l = 0}^K \lambda^l).$$
Take $f_K$ to be the unique value that satisfies this equation.
Once we have defined $f_K$, again apply the constraint and
take $f_{{K - 1}}$
to be the unique value that
satisfies
\begin{align*}
d(K-1)
&= d_{\lambda}(K,K-1) + d_{\lambda}({K-1},K-1)\\
&= f_{K} \lambda^{K-1} / (\sum_{l = 0}^K
\lambda^l ) + f_{{K-1}} \lambda^{K-1}
/ (\sum_{l = 0}^{K-1} \lambda^l).
\end{align*}
Iterating this process uniquely defines
$\vec{f}$ as the unique value that satisfies
$$d(j) = \sum_{i = j}^K d_{\lambda}(i,j) =
\sum_{i = j}^{K} f_{i} \lambda^{j} / (\sum_{l = 0}^i \lambda^l),$$
or
$$f_{i} = (\sum_{l = 0}^{i} \lambda^l) / \lambda^i
\left( d(i) - \sum_{j = i+1}^{K}
f_{j} \lambda^{i} / (\sum_{l = 0}^j \lambda^l) \right).$$

However, $\vec{f}$ may not be an explanation, since
some $f_j$ may be negative.
This happens exactly when
\begin{equation}\label{eq:explanation}
d(i) < \sum_{j = i+1}^K f_{j} \lambda^i / (\sum_{l = 0}^j
\lambda^l).
\end{equation}
As $\lambda$ grows large, the right-hand side
of~(\ref{eq:explanation})
tends to 0.  Since $d$ is fully-supported, we must have $d(i) >
0$.  Thus, we can ensure that (\ref{eq:explanation}) does not hold for any
$i$ by taking $\lambda$ sufficiently large.
Thus, for sufficiently large $\lambda$, $\vec{f}$
provides an explanation for $d$.
Continuing to increase $\lambda$ will give an infinite number of
different explanations.
\eprf

\bibliographystyle{spbasic}
\bibliography{Z:/Research/Bibliography/kash}

\end{document}